# Electronic Structure, Optical Response, Thermal and Mechanical Behavior of $B_6X$ (X = S, Se) under Pressure: A Comprehensive Ab-initio Exploration

Sourav Kumar Sutradhar, Tanvir Khan, Tauhidur Rahman, Sraboni Saha Moly, Mst. Maskura Khatun, S.H. Naqib*

Department of Physics, University of Rajshahi, Rajshahi 6205, Bangladesh

*Corresponding Author; Email: salehnaqib@yahoo.com

## Abstract

This study presents a comprehensive first-principles investigation of the pressure-dependent structural, electronic, optical, mechanical, and bonding properties of orthorhombic boron-rich chalcogenides $B_6X$ (X = S, Se). Calculations were performed using density functional theory across a wide range of hydrostatic pressures (0 – 20 GPa). The computed elastic constants, bulk, Young, shear moduli, and Poisson's ratio, revealed mechanical robustness and strong resistance to deformation, even under significant compression. Electronic band structure and density of states (DOS) analyses indicate that the materials exhibit indirect bandgap semiconducting behavior. The band gaps are notably sensitive to pressure, suggesting that they can be tuned for high-performance photovoltaic and optoelectronic applications. Optical parameters. These results reveal clear pressure-induced spectral shifts, particularly in the visible and ultraviolet regions, suggesting modified light-matter interaction under compression. Bonding characteristics were explored through Mulliken population analysis and electron density difference maps. These analyses confirm the mixed covalent-ionic nature of the B–X bonds, with covalency levels influenced by the chalcogen atoms and external pressure. Phonon dispersion curves verified the dynamical stability of both materials within the investigated pressure range. Hardness estimations, combined with elastic parameters, and melting temperatures, further indicate that $B_6S$ and $B_6Se$ possess significant mechanical strength suitable for applications under harsh environments. The thermal properties suggest that both these compounds possess features suitable to be used as excellent thermal barrier coating materials. The integrated analysis of diverse set of physical properties offers a detailed understanding of the pressure-property relationship in $B_6X$ (X = S, Se) compounds. These findings provide a theoretical framework to guide future experimental synthesis and optimization, highlighting the potential of $B_6S$ and $B_6Se$ optoelectronic, thermal, and mechanically robust applications.

## 1. Introduction

The advancement of modern industry requires materials with exceptional physical and chemical properties to meet emerging energy and technological demands [1-13].

Unconventional crystal structures often exhibit unique characteristics that play a vital role in advanced engineering and smart functional materials science. Boron-rich systems, for instance, are based on $B_{12}$ closed-cage clusters (a three-dimensional rigid network) [14,15], which are relatively uncommon. Boron and its compounds containing $B_{12}$ icosahedral units attract both fundamental and technological interest due to their outstanding properties, including high hardness with good mechanical strength, low density, extremely high melting points, and excellent thermal and chemical stability. These features make them highly promising for applications in high-temperature and high-pressure devices [16,17]. Specifically, boron-rich compounds are widely used in heavy-duty tools that operate in abrasive environments and under high strain, such as turbine and rotor blades, sports equipment, advanced biological and chemical sensors, and protective armor materials [17,18]. In 2020, boron-rich chalcogenide compounds $B_6X$ (X = S, Se) with $B_{12}$ closo-cluster structures were synthesized at 6.1 GPa and 2700 K in a toroid-type high-pressure apparatus, and their stability was further supported by formation enthalpy and phonon calculations using the Vienna *ab initio* Simulation Package (VASP). According to recent experimental findings, the structural stability of $B_6X$ (X = S, Se) compounds varies under hydrostatic pressure; while $B_6S$ remains stable throughout the 0-20 GPa range, $B_6Se$ stabilizes under a narrower pressure window of 5-20 GPa [19]. The study confirmed that these compounds adopt an orthorhombic crystal structure with space group *Pmna*. The mechanical properties, including polycrystalline elastic moduli and hardness, have been predicted for this class of materials, identifying them as members of the family of hard phases with a hardness of approximately 31 GPa [6]. Several boron sulfide compounds, such as r-BS, $B_2S_3$ (II), and $B_2S_3$ (III), with different crystal symmetry groups, have also been synthesized under high-pressure and high-temperature conditions [16,17,20]. Zhang *et al*. [21] investigated the phase stability, mechanical, and electronic properties of the Si-B system and related phases, including α-$SiB_3$ ($SiB_3$, $SiB_4$), β-$SiB_3$, $SiB_6$, and $SiB_{36}$, using first-principles calculations. Their study reported bulk moduli in the range of 120-180 GPa and shear moduli between 55-154 GPa. In another work, boron-rich solids $B_6S_{1-x}$ ($0.37 < x \leq 0.40$) with a rhombohedral structure based on a $B_{12}$ icosahedral framework were shown to exhibit semiconducting behavior with a maximum Seebeck coefficient of 220 $\mu V\ K^{-1}$ [22]. Other compounds, such as hexacarbides and silicides, have also been reported. More recently, a metastable boron-rich carbide $B_6C$ was identified, displaying both superhard and superconducting properties [23]. Yuan *et al*. [24] predicted a novel metallic silicon hexaboride, $B_6Si$, which crystallizes in the *Cmca* space group. The structure consists of three-dimensional networks of $B_{12}$ icosahedra interconnected with silicon atoms.

However, it is important to note that existing investigations on boron-rich compounds of this type have so far been restricted mainly to structural, Raman spectroscopy, and basic mechanical properties such as elastic moduli and hardness. A significant knowledge gap remains in understanding other critical aspects on varying pressure, including electronic band structure, optical, thermal, and additional mechanical characteristics such as elastic stiffness constants, fracture toughness, machinability index, Cauchy pressure, anisotropy, brittleness/ductility, and Poisson's ratio. Pressure is a clean tunning parameter that does not introduce defects and disorders which are unavoidable in case of doping. Moreover, pressure induced changes in the ground state physical properties give useful scientific insights regarding the system. It is also useful to optimize the possible applications under harsh physical conditions. These insights are essential to fully exploit the potential of $B_6X$ (X = S, Se) compounds for practical applications. In particular, knowledge of the electronic features (band structure, electronic density of states, charge density distribution, Mulliken population analysis, and electron localization function) and optical behavior (absorption coefficient, reflectivity, photoconductivity, refractive index, loss function, and dielectric response) under varying pressure 0 GPa to 20 GPa is crucial for designing structural, photovoltaic, and optoelectronic devices. Likewise, understanding mechanical anisotropy is of great importance, as it strongly influences key behaviors such as crack initiation and propagation, crystal instability, phase transformation, and plastic deformation factors that often limit the practical use of materials. Furthermore, insight into thermodynamic properties such as Debye temperature, temperature-dependent lattice thermal conductivity, thermal expansion coefficients, and specific heat, along with their correlations to mechanical performance, could provide a new perspective for employing these compounds in devices operating under extreme conditions. In this work we wish to bridge the existing research gaps and to present a comprehensive understanding of the physical properties of $B_6X$ (X = S, Se) compounds under varying pressure conditions.

## 2. Methodology

Density functional theory (DFT), which solves the Kohn-Sham equation under periodic boundary conditions (including Bloch states), is one of the most widely applied approaches for *ab initio* studies of crystalline materials. In this work, the CASTEP code, based on DFT, was employed to investigate the physical properties of the compound [10,11]. This code uses the plane-wave pseudopotential method for total energy calculations. Several exchange-correlation functionals were tested in the present study, including the Perdew-Burke-Ernzerhof (PBE) Generalized Gradient Approximation (GGA) [12], the Perdew-W91 GGA [13], the RPBE GGA [14], the PBEsol GGA [12], and the Local Density Approximation (LDA) [15,16]. Accurate structural optimization plays a key role, and the GGA-PBE provided the most reliable structural parameters for $B_6X$ (X= S, Se). To describe electron-ion interactions, Vanderbilt-type ultrasoft pseudopotentials were applied [17]. This approach offers computational efficiency.

The Brillouin zone (BZ) sampling was carried out using the Monkhorst and Pack scheme [18] with a mesh size of 15 × 17 × 11 for $B_6S$ and 12 × 13 × 8 for $B_6Se$ compound was used. A plane-wave cutoff energy of 550 eV was chosen. Additional details regarding pseudopotentials and k-point grids can be found in [19]. The stress-strain method was employed to calculate the single-crystal elastic constants ($C_{ij}$) of the crystal structure [20]. For the optimized structure, macroscopic elastic properties such as the bulk modulus (B), shear modulus (G), and Young's modulus were obtained through the Voigt-Reuss-Hill (VRH) averaging scheme [21,22].

The optimized geometry of $B_6X$ (X = S, Se) was further used to compute the electronic band structure, partial density of states (PDOS), and total density of states (TDOS). Optical properties were derived from the complex dielectric function, expressed as $\varepsilon(\omega) = \varepsilon_1(\omega) + i\varepsilon_2(\omega)$. The real component, $\varepsilon_1(\omega)$, was calculated from the imaginary part, $\varepsilon_2(\omega)$, using the Kramers-Kronig relation [23]. The imaginary part itself was determined from the momentum matrix elements of photon-induced transitions between occupied and unoccupied electronic states, following the expression for $\varepsilon_2(\omega)$ implemented in CASTEP:

$$\varepsilon_2(\omega) = \frac{2e^2\pi}{\Omega\varepsilon_0}\sum_{kvc}|< \Psi_k^c|\hat{u}.\vec{r}|\Psi_k^v >|^2\,\delta(E_k^c - E_K^v - E) \quad (1)$$

The unit cell volume is represented by **Ω**, the angular frequency (or energy) of the incident photon by ω, and the unit vector û specifies the polarization direction of the electric field. The electronic charge is denoted as *e*, while $\Psi_k^c$ and $\Psi_k^v$ correspond to the conduction and valence band wave functions at a given wave vector *k*, respectively. The delta function in Equation (1) ensures that both energy and momentum conservation are maintained during optical transitions. Once the dielectric function ε(ω) is determined over a range of energies/frequencies, all other key optical characteristics can be derived using the relationships provided in [24,25]. In particular, the real part $n(\omega)$ and the imaginary part $k(\omega)$ of the complex refractive index can be obtained through the following expressions:

$$n(\omega) = \frac{1}{\sqrt{2}}\left[\{\varepsilon_1(\omega)^2 + \varepsilon_2(\omega)^2\}^{\frac{1}{2}} + \varepsilon_1(\omega)\right]^{\frac{1}{2}} \quad (2)$$

$$k(\omega) = \frac{1}{\sqrt{2}}\left[\{\varepsilon_1(\omega)^2 + \varepsilon_2(\omega)^2\}^{\frac{1}{2}} - \varepsilon_1(\omega)\right]^{\frac{1}{2}} \quad (3)$$

Furthermore, the reflectivity $R(\omega)$ can be determined directly from the real and imaginary components of the complex refractive index:

$$\mathrm{R}(\omega) = \frac{(n-1)^2 + k^2}{(n+1)^2 + k^2} \quad (4)$$

The absorption coefficient α(ω), optical conductivity σ(ω), and energy loss function L(ω) can be evaluated using the following relations:

$$\alpha(\omega) = \frac{4\pi k(\omega)}{\lambda} \quad (5)$$

$$\Sigma(\omega) = \frac{2W_{cv}\hbar\omega}{\vec{E}_0^2} \quad (6)$$

$$\mathrm{L}(\omega) = \mathrm{Im}\left(-\frac{1}{\varepsilon(\omega)}\right) \quad (7)$$

In Equation (6), $W_{cv}$ represents the optical transition probability per unit time.

To gain insight into the bonding characteristics of solids, Mulliken bond population analysis and charge density distribution were employed. In the case of $B_6X$ (X = S, Se), the method involves projecting the plane-wave states onto a linear combination of atomic orbital (LCAO) basis sets [26].

In this work, the third-order Birch-Murnaghan equation of state [27] was employed to generate energy-volume data based on zero-temperature and zero-pressure equilibrium values for energy, volume, and bulk modulus, as obtained from density functional theory (DFT) calculations. Additionally, the bond hardness of the covalent crystal was estimated using a widely adopted semiempirical formula. This approach enables a comprehensive analysis of the thermomechanical, elastic, and structural properties as they change under varying pressures. By applying the Birch-Murnaghan equation, the study establishes the relationship between energy and volume under different pressure conditions, offering valuable insights into the evolution of the material's properties during compression.

## 3. Results and Discussion

### 3.1 Structural Properties

The physical and electrical properties of a solid are largely influenced by how its atoms are arranged in the unit cell. The $B_6X$ compounds (X = S or Se) adopt an orthorhombic crystal structure characterized by three unique lattice parameters: a, b, and c. In this structure, the atoms B, S, and Se occupy the 4c Wyckoff positions within the *Pmna* space group (No. 53) [28,29]. Each unit cell comprises 28 atoms, including 24 boron atoms and 4 chalcogen atoms (S or Se) positioned at a single independent 4h site, and contains four formula units (Z = 4). The associated point group of this structure is *Pmna*. The crystal structure of the $B_6X$ compounds (X = S, Se) can be effectively described using a trigonal prismatic arrangement. Each unit cell contains 28 atoms, comprising 24 boron atoms arranged in $B_{12}$ icosahedra that occupy four independent Wyckoff positions (e.g., 4h and 8i), and 4 chalcogen atoms (S or Se) located at a single independent 4h site [28]. The optimized lattice parameters for $B_6S$ and $B_6Se$, along with values from previous studies, are summarized in **Table 1**. This table provides a comparison between the optimized parameters and available experimental and theoretical data. According to **Table 1**, the lattice constants a, b, c, and the unit cell volume V, calculated using the GGA (PBE) functional [30], show deviations of no more than 0.13% for $B_6S$ and 0.17% for $B_6Se$ compared to experimental measurements, indicating excellent agreement. The atomic coordinates are given in **Table 2**.

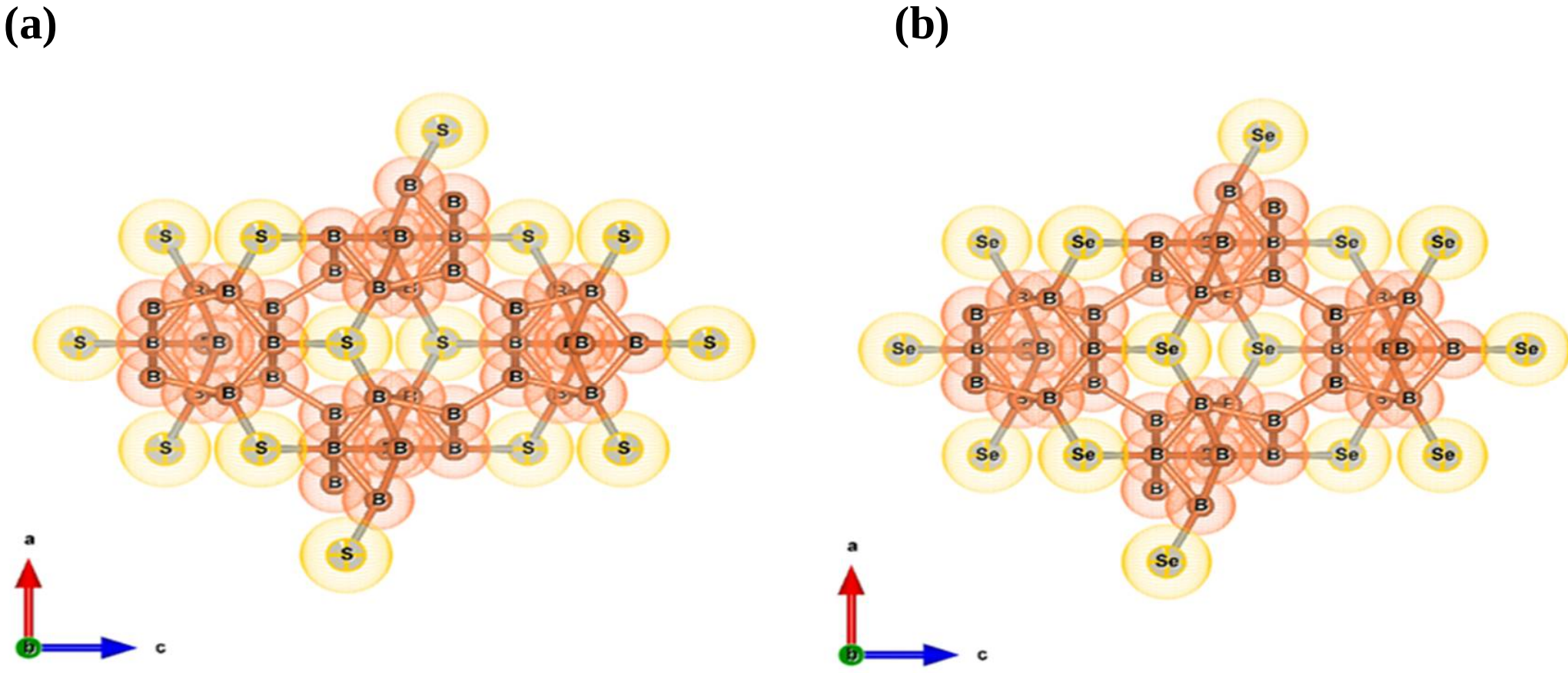

**Figure 1:** Schematic crystal structure of (a) $B_6S$ and (b) $B_6Se$. The crystallographic directions are shown.

**Table 1:** Calculated and previously obtained lattice constants (a, b, and c all in Å) and equilibrium volume (V in Å$^3$) of the $B_6X$ (X = S, Se) compounds.

| Compound | P (GPa) | a | b | c | V | Ref. |
|---|---|---|---|---|---|---|
| $B_6S$ | - | 5.82 | 5.30 | 8.21 | 253.34 | [28] |
| | 0 | 5.82 | 5.31 | 8.19 | 252.99 | This work |
| | 5 | 5.76 | 5.22 | 8.12 | 244.54 | |
| | 10 | 5.71 | 5.15 | 8.08 | 237.49 | |
| | 15 | 5.71 | 5.15 | 8.08 | 237.41 | |
| | 20 | 5.61 | 5.03 | 7.99 | 225.33 | |
| $B_6Se$ | - | 5.95 | 5.36 | 8.38 | 267.06 | [28] |
| | 5 | 5.87 | 5.28 | 8.28 | 256.81 | This work |
| | 10 | 5.80 | 5.19 | 8.23 | 247.47 | |
| | 15 | 5.75 | 5.14 | 8.18 | 241.51 | |
| | 20 | 5.70 | 5.08 | 8.13 | 235.70 | |

**(a)** **(b)**

**Figure 2:** Normalized lattice parameters $a/a_0$, $b/b_0$, $c/c_0$, and $V/V_0$ for (a) $B_6S$ and (b) $B_6Se$ under different pressures.

**Table 2**: Fractional coordinates of the atoms in the unit cell of $B_6X$ (X = S, Se) compounds.

| Compound | Atom | Fractional coordinates | | | Ref. |
|---|---|---|---|---|---|
| | | *x* | *y* | *z* | |
| $B_6X$ (X = S, Se) | S1 (4h) | 0.000 | 0.186 | 0.868 | [28] |
| | B1 (4h) | 0.000 | 0.157 | 0.483 | |
| | B2 (8i) | 0.838 | 0.356 | 0.335 | |
| | B3 (4h) | 0.000 | 0.345 | 0.666 | |
| | B4 (8i) | 0.259 | 0.351 | -0.043 | |
| | Se1 (4h) | 0.000 | 0.169 | 0.866 | |
| | B1 (4h) | 0.000 | 0.160 | 0.480 | |

| | | | | |
|---|---|---|---|---|
| | B2 (8i) | 0.840 | 0.363 | 0.336 |
| | B3 (4h) | 0.000 | 0.349 | 0.662 |
| | B4 (8i) | 0.263 | 0.3531 | -0.039 |

The pressure dependent variations of the structural parameters are shown in **Figure 1**. It should be noted that all the results for $B_6Se$ are presented within the pressure range of 5 – 20 GPa, taking account of its range of structural and dynamical stability [19].

**3.2 Elastic and Mechanical Properties**

The technology of materials heavily rely on elastic characteristics. These characteristics provide crucial information about the nature of the forces acting in solids and connect the mechanical and dynamical behavior of solids. Specifically, they include details on the stiffness and stability of materials [31,32]. To determine the orthorhombic $B_6S$ and $B_6Se$ compound's elastic constants, nine separate strains are required. **Table 3** displays the elastic constants computed at various pressures. The mechanical stability criteria for orthorhombic crystals under hydrostatic pressure is expressed as follows [33]:

$$C_{44} - P > 0; C_{55} - P > 0; C_{66} - P > 0; C_{11} - P > 0;$$

$$(C_{11} - P)(C_{22} - P) > (C_{12} + P)^2$$

$$\begin{aligned} &[C_{11}C_{22}C_{33} + 2C_{12}C_{13}C_{23} - C_{11}C_{23}^2 - C_{22}C_{13}^2 - C_{33}C_{12}^2] \\ &-P\begin{bmatrix} C_{11}C_{22} + C_{22}C_{33} + C_{11}C_{33} - C_{12}^2 - C_{13}^2 - C_{23}^2 - \\ 2(C_{12}C_{13} + C_{12}C_{23} + C_{13}C_{23}) + 2(C_{11}C_{23} + C_{22}C_{13} + C_{33}C_{12}) \end{bmatrix} \\ &+P^2[C_{11} + C_{22} + C_{33} + 2(C_{12} + C_{13} + C_{23})] - 4P^3 > 0 \end{aligned} \quad (8)$$

**Table 3** Single crystal elastic constants ($C_{ij}$ in GPa), Cauchy pressure (CP in GPa), The machinability index ($\mu^M$) of $B_6X$ (X = S, Se) compounds.

| Compound | P (GPa) | $C_{11}$ | $C_{22}$ | $C_{33}$ | $C_{44}$ | $C_{55}$ | $C_{66}$ | $C_{12}$ | $C_{13}$ | $C_{23}$ | CP | $\mu^M$ | Ref. |
|---|---|---|---|---|---|---|---|---|---|---|---|---|---|
| $B_6S$ | 0 | 417 | 277 | 463 | 115 | 203 | 78 | 2 | 75 | 32 | - | - | [28] |
| | 0 | 419.84 | 284.59 | 457.51 | 119.17 | 203.21 | 88.53 | 7.36 | 69.94 | 30.87 | -111.8 | 1.24 | This work |
| | 5 | 441.29 | 313.87 | 498.32 | 127.21 | 215.96 | 90.96 | 13.95 | 88.60 | 42.41 | -113.3 | 1.30 | |
| | 10 | 454.07 | 328.12 | 527.59 | 131.02 | 222.61 | 77.19 | 14.23 | 107.87 | 57.30 | -116.8 | 1.36 | |
| | 15 | 477.56 | 366.54 | 565.45 | 146.91 | 233.21 | 99.77 | 28.32 | 115.95 | 66.13 | -118.6 | 1.34 | |
| | 20 | 494.18 | 397.05 | 597.08 | 157.24 | 242.33 | 108.48 | 41.79 | 129.33 | 79.59 | -115.5 | 1.37 | |
| $B_6Se$ | 0 | 377 | 275 | 450 | 137 | 182 | 90 | 5 | 60 | 48 | - | - | [28] |
| | 5 | 411.33 | 305.39 | 482.47 | 148.20 | 197.59 | 134.63 | 18.25 | 67.75 | 49.36 | -129.9 | 1.07 | This work |
| | 10 | 459.38 | 384.01 | 533.15 | 181.37 | 208.46 | 164.74 | 60.70 | 84.36 | 77.05 | -120.7 | 1.10 | |
| | 15 | 472.18 | 391.48 | 564.07 | 183.49 | 218.11 | 169.24 | 58.82 | 93.37 | 83.69 | -124.7 | 1.13 | |
| | 20 | 484.16 | 425.50 | 591.96 | 195.33 | 225.97 | 171.33 | 73.60 | 109.78 | 96.58 | -121.7 | 1.16 | |

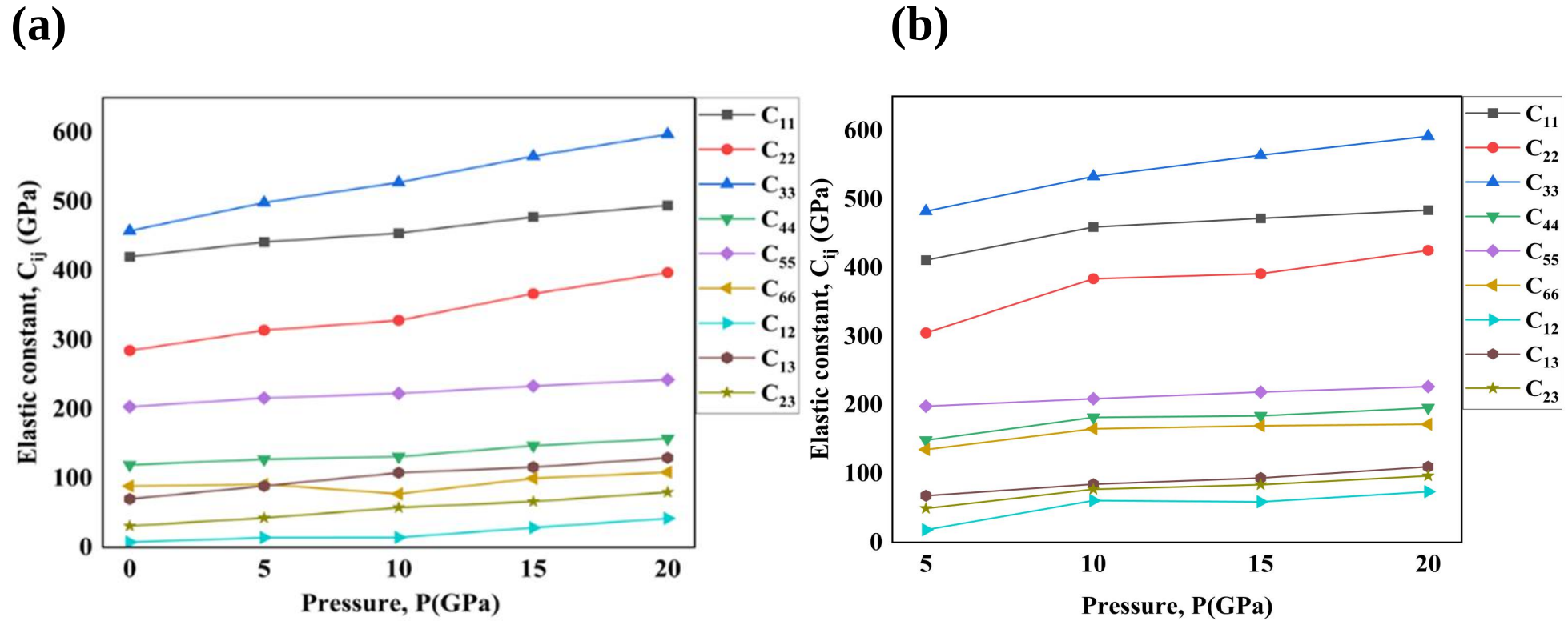

**Figure 3:** Elastic constants for (a) $B_6S$ and (b) $B_6Se$ under different pressures.

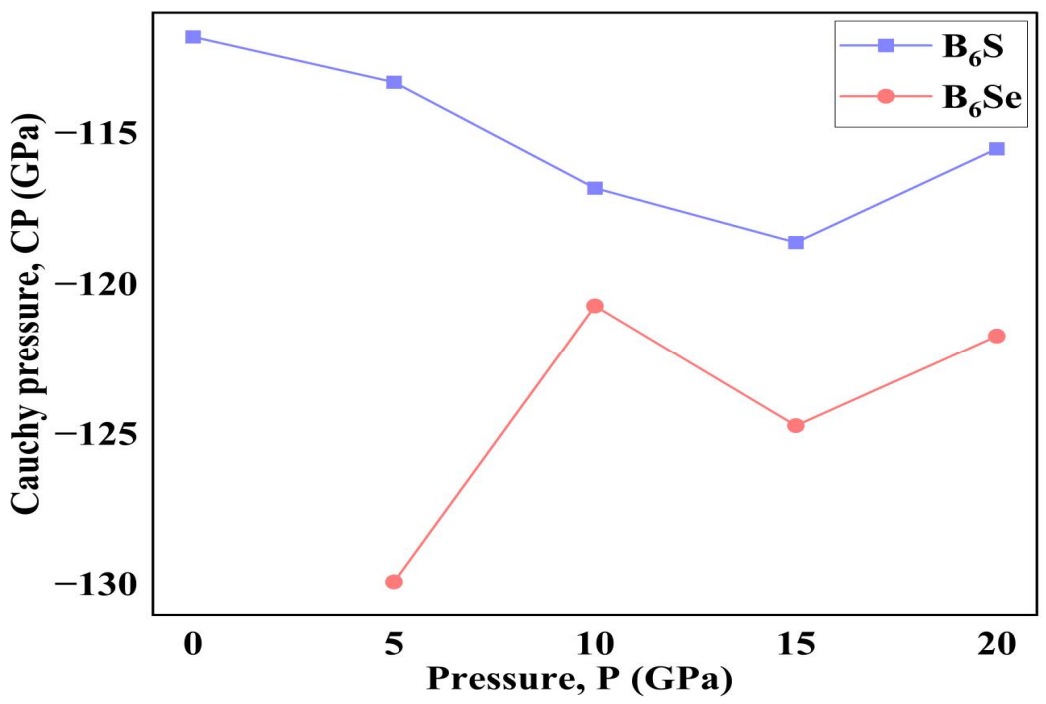

**Figure 4:** Cauchy pressure for $B_6S$ and $B_6Se$ under different pressures.

As can be seen, every $C_{ij}$ value is positive, satisfying elastic stability conditions. As a result, ternary boride compounds $B_6X$ (X = S, Se) have mechanical stability within the pressure range considered. Furthermore, it is evident that the ground state elastic constants determined here are in good agreement with the previously calculated values [29]. Among the nine independent elastic constants, the diagonal elastic constants $C_{11}$, $C_{22}$, and $C_{33}$ measure the resistance to linear compression along the a-, b-, and c-axes of orthorhombic crystals, respectively. The calculated values of $C_{22}$ for $B_6S$ and $B_6Se$ are less than $C_{11}$ and $C_{33}$, suggesting that these compounds are more compressible along the b-axis than along the a- and c-axes. This results from the differences in atomic arrangements and strength of atomic bondings along different crystallographic axes. The fact that $C_{33} > (C_{11}, C_{22})$ for both compounds indicates that the atomic bonding strengths in these hard phase compounds are greatest in the c-direction. The shear and off-diagonal components of the elastic constants are $C_{12}$, $C_{13}$, and $C_{23}$. The most important factor influencing a material's indentation hardness is its elastic constant $C_{44}$. The materials' significant resistance to shear deformation in the (100) plane is shown by the large value of $C_{44}$, whereas the resistance to shear in the <011> and <110> directions is indicated by $C_{55}$ and $C_{66}$, respectively. It is discovered that $B_6Se$ has a larger elastic constant $C_{44}$. $B_6Se$ should be able to resist strongly shape changing stresses and should have greater level of hardness compared to $B_6S$. The pressure dependent variations in the elastic constants are shown in **Figure 3**. All the elastic constants increase with rising pressure, though the rates are different. This behavior is standard for crystalline solids.

The Voigt-Reuss-Hill (VRH) approximation [34-36] is a widely used method in materials science for estimating the effective elastic moduli [bulk (B), shear (G), and Young's (Y) moduli] of polycrystalline aggregates from the elastic constants of their constituent single crystals. This approach combines the Voigt bound, which assumes uniform strain throughout the aggregate and provides an upper limit by averaging the elastic stiffness constants, and the Reuss bound, which considers uniform stress and provides a lower limit by averaging the elastic compliances. The Hill approximation simply takes the arithmetic mean of these two bounds, resulting in a value that often closely matches the experimental results for solids.

The physical significance of the three elastic moduli derived from this averaging lies in their description of a material's resistance to different types of deformation. The bulk modulus quantifies a material's resistance to uniform, isotropic compression, directly relating an applied hydrostatic pressure to the resulting fractional volume change. It is fundamentally a measure of the energy required to alter the bond lengths without changing the bond angles. The shear modulus, on the other hand, describes resistance to shape change at constant volume, representing the material's rigidity under shearing forces and it reflects the energy needed to distort the bond angles. Young's modulus, the most common measure of stiffness, quantifies the uniaxial tensile or compressive resistance. This parameter couples both volume and shape changes into a single ratio of uniaxial stress to resulting normal strain. Together, these elastic moduli provide a complete picture of a material's elastic behavior, from its compressibility and rigidity to its overall stiffness. The computed values of these elastic moduli are presented in **Table 4** at different pressures. The subscripts 'V' and 'R' signify the Voight and Reuss bounds. B, G, and Y are the values obtained using Hill's approximation. The details regarding these calculations can be found elsewhere [37-40].

The ease of machining a material is dictated by its machinability index ($\mu^{M}$). This index can be used to determine the ideal cutting parameters for a given material. Furthermore, a material's flexibility and dry lubricating qualities can be examined using this index [40]. Similar to numerous well-known hard phase materials [41,42], the $B_6X$ (X = S, Se) compounds' $\mu^{M}$ values (see **Table 3**) implies a moderate level of machinability. The following formula has been used to determine the machinability index ($\mu^{M}$):

$$\mu^{M} = \frac{B}{C_{44}} \tag{9}$$

Higher bulk modulus alongside decreased shear resistance results in improved machinability and dry lubricity.

Materials can be categorized as either brittle or ductile using indicators such as Cauchy pressure (CP), Poisson's ratio (ν), and Pugh's ratio (B/G). The nature of atomic bonding within a material is also reflected in its Cauchy pressure [43,44]. A positive CP is generally associated with ductile behavior, while a negative CP value suggests brittleness [45]. According to Pettifor [46], large positive CP values correspond to strong metallic, non-directional bonding, whereas negative values are linked to angular bonding. When bonding can be modeled by simple pair potentials, such as the Lennard-Jones potential, the Cauchy pressure becomes zero. The Cauchy pressure is calculated using the relation:

$$\mathrm{CP} = (C_{23} - C_{44}) \tag{10}$$

The pressure dependent CP is shown in **Figure 4**. The behavior is nonmonotonic. This arises from the fact that the changes in $C_{23}$ and $C_{44}$ with pressure have different gradients at different pressures.

The Poisson's ratio (ν) is widely employed to evaluate a material's ductility or brittleness, as well as its bonding characteristics. It also reflects the compound's stability under shear. Lower values indicate greater resistance to shear deformation. Typically, ν lies within the range of $-1.0 \leq \nu \leq 0.50$ [31,32]. When $\nu < 0.26$, the material is considered brittle, whereas $\nu > 0.26$ signifies ductile behavior.

The Poisson's ratio was determined using the following relation [47,48]:

$$\nu = \frac{(3B - 2G)}{2(3B + G)} \tag{11}$$

Pugh's ratio (B/G), defined as the ratio of the bulk modulus to the shear modulus, is another important criterion for distinguishing between brittle and ductile behavior in solids [47,49,50]. The critical threshold separating the two lies at approximately 1.75. A Pugh's ratio below 1.75 suggests brittleness, while values above this limit indicate ductility [50]. All the indicators suggest that the compounds under study remains brittle within the pressure range considered in this paper. The values of Poisson's ratio and Pugh's ratio are presented in **Table 4**. Both Pugh's ratio and Poisson's ratio increase systematically with increasing pressure.

Hardness, defined as a material's resistance to permanent deformation, is one of the most critical mechanical properties ensuring performance in industrial applications. It plays a key role in areas such as tribological coatings, heavy-duty components, low-emission window glass coatings, optoelectronic devices, and microelectronics [51-53]. Understanding the connection between hardness and other material characteristics such as chemical stability, scratch resistance, and surface durability is essential. Hard materials are widely employed in the production of cutting tools and wear-resistant coatings, owing to the strong correlation between hardness and strength [54]. Several theoretical models have been developed to estimate hardness in relation to parameters like Poisson's ratio (ν), Young's modulus (Y), bulk modulus (B), and shear modulus (G). Notable frameworks in this field have been proposed by N. Miao et al. [55,56], X. Chen et al. [57,58], Y. Tian et al. [59], D. M. Teter [60], and E. Mazhnik et al. [61]. Variations of elastic moduli and other elastic descriptors with pressure are illustrated in **Figures 5 and**

**6**. All three bulk elastic moduli increase with increasing pressure (**Figure 5**). This is a standard pressure response found in solids with different crystal symmetry.

**Table 4:** Polycrystalline bulk modulus ($B_V$, $B_R$, $B$ in GPa), shear moduli ($G_V$, $G_R$, $G$ in GPa), Young modulus ($Y$ in GPa), Pugh's ratio ($B/G$) and Poisson's ratio (n), tetragonal shear modulus, ($C_t$ in GPa), Vickers hardness ($H_V$ in GPa) for the $B_6X$ (X= S, Se) compounds in the ground state.

| Compound | P (GPa) | $B_V$ | $B_R$ | B | $G_V$ | $G_R$ | G | Y | $C_t$ | (B/G) | $v$ | $H_V$ | Ref. |
|---|---|---|---|---|---|---|---|---|---|---|---|---|---|
| $B_6S$ | 0 | 153.14 | 143.23 | 148.19 | 152.43 | 137.22 | 144.82 | 327.71 | 206.24 | 1.02 | 0.13 | 30.36 | |
| | 5 | 171.49 | 160.73 | 166.11 | 160.73 | 144.12 | 152.42 | 350.16 | 213.67 | 1.09 | 0.15 | 29.30 | |
| | 10 | 185.40 | 172.31 | 178.85 | 161.52 | 138.46 | 149.99 | 351.67 | 219.92 | 1.19 | 0.16 | 26.15 | |
| | 15 | 203.37 | 191.88 | 197.63 | 175.92 | 159.09 | 167.51 | 391.82 | 224.38 | 1.18 | 0.17 | 28.63 | |
| | 20 | 221.07 | 210.06 | 215.57 | 184.12 | 168.60 | 176.36 | 415.71 | 226.20 | 1.22 | 0.18 | 28.52 | This work |
| $B_6Se$ | 5 | 163.33 | 154.93 | 159.13 | 167.00 | 161.86 | 164.43 | 366.91 | 196.54 | 0.97 | 0.12 | 35.39 | |
| | 10 | 202.31 | 197.95 | 200.12 | 187.88 | 185.51 | 186.69 | 427.23 | 199.34 | 1.07 | 0.14 | 34.45 | |
| | 15 | 211.05 | 205.00 | 208.02 | 193.63 | 190.69 | 192.16 | 440.77 | 206.68 | 1.08 | 0.15 | 34.79 | |
| | 20 | 229.06 | 223.47 | 226.26 | 199.97 | 197.09 | 198.52 | 460.81 | 205.28 | 1.14 | 0.16 | 33.58 | |

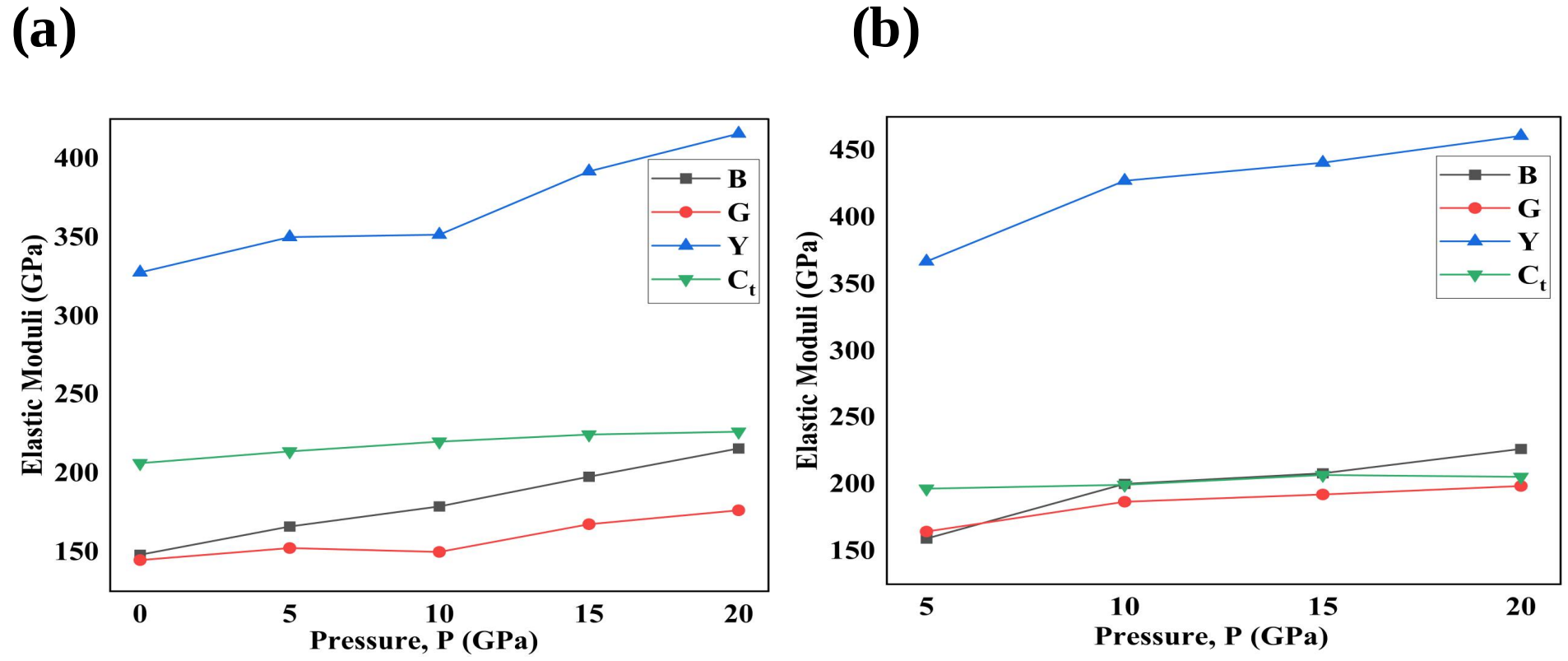

**Figure 5:** Elastic moduli for (a) $B_6S$ and (b) $B_6Se$ under different pressures.

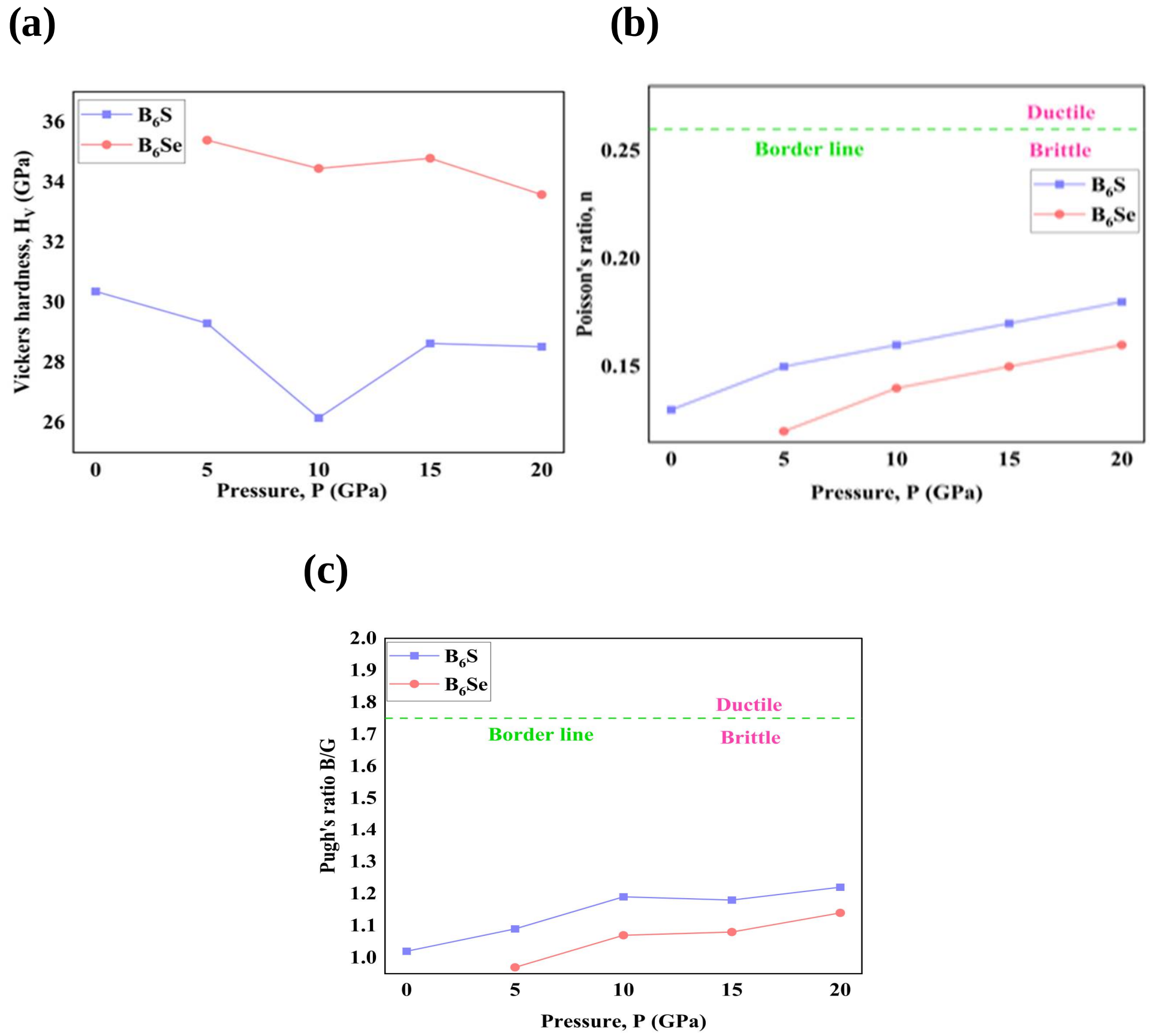

**Figure 6:** (a) Vickers hardness (b) Poisson's ratio and (c) Pugh's ratio for $B_6S$ and $B_6Se$ under different pressure.

In **Figures 6(b)** and **(c),** the Poisson's ratio and Pugh's ratio vary with increasing pressure and take an uptrend with increasing pressure. All the relevant parameters indicate the brittleness of $B_6S$ and $B_6Se$ compounds. The Vicker's hardness exhibits nonmonotonic variation with pressure [**Figure 6(a)**]. Irrespective of pressure, both the compounds show fairly hard characteristics. $B_6S$ is substantially harder than $B_6Se$ at all pressures considered.

The Kleinman parameter (ζ) provides a qualitative measure of the relative contributions of bond bending and bond stretching to a material's mechanical strength. It can also be used to describe the relative displacement of cation and anion sublattices under volume-conserving distortions, where atomic positions are not strictly constrained by crystal symmetry [62,63]. The Kleinman parameter is calculated using the following relation:

$$\zeta = \frac{C_{11} + 8C_{12}}{7C_{11} + 2C_{12}} \tag{12}$$

**Table 5** presents the calculated Kleinman parameters for $B_6S$ and $B_6Se$. As shown, the ζ value for $B_6S$ is below 0.5, while $B_6Se$ exhibits a similar value. This indicates that, in both compounds, the mechanical strength is predominantly governed by bond stretching.

**Table** Error! No text of specified style in document.**:** Kleinman parameter (ζ) of $B_6X$ (X = S, Se) compounds.

| Compound | P (GPa) | ζ | Ref. |
|---|---|---|---|
| $B_6S$ | 0 | 0.16 | This work |
| | 5 | 0.17 | |
| | 10 | 0.18 | |
| | 15 | 0.21 | |
| | 20 | 0.23 | |
| $B_6S$ | 5 | 0.19 | |
| | 10 | 0.27 | |
| | 15 | 0.28 | |
| | 20 | 0.30 | |

### 3.3 Elastic Anisotropy

Unlike cubic crystals, the bulk modulus of anisotropic crystals exhibits directional dependence. In such cases, elastic anisotropy arises not only from shear anisotropy but also from the anisotropy of the linear bulk modulus, meaning that any single anisotropy factor alone cannot fully capture the elastic behavior [64].

Mechanical anisotropy plays a crucial role in determining a material's mechanical stability and the way it responds to structural stresses under different loading conditions. There are a large number of anisotropy indices signifying anisotropy in different elastic moduli and parameters [65-69].The elastic anisotropy of crystals can be effectively illustrated through two- and three-dimensional (2D and 3D) graphical representations of key elastic parameters, such as Young's modulus (Y), linear compressibility (β), shear modulus (G), and Poisson's ratio (***ν***). The elastic stiffness matrices for the $B_6S$ and $B_6Se$ compounds were calculated using the CASTEP code, and these matrices were subsequently processed with the ELATE program [70] to directly visualize the anisotropy levels of Young's modulus, linear compressibility, shear modulus, and Poisson's ratio for the $B_6S$ and $B_6Se$ compounds. In these visualizations, spherical (3D) and circular (2D) shapes represent the isotropic nature of crystals, while deviations from these shapes indicate an increasing degree of anisotropy. **Figures 7-10** for $B_6S$ and **11-14** for $B_6Se$ show the directional dependence of the elastic parameters for the $B_6S$ and $B_6Se$ compounds in both three- and two-dimensional forms, plotted along the (xy)ab, (xz)ac, and (yz)bc planes, respectively. The elastic behaviors are shown for 0 and 20 GPa. The plots show systematic pressure dependence over the entire pressure range considered. The essential features of this analyses are tabulated in **Table 6**. The anisotropy indicator $A_j$ ($j$ = Y, β, G, ν) is determined from the ratio of the maximum value and the minimum value of the respective elastic parameter given in **Table 6**. This found that pressure strongly modifies all the anisotropy indicators. The behavior is nonmonotonic. This is because the atomic bondings along different directions respond to pressure differently. It becomes clear from **Figures 7-14** and the data presented in **Table 6** that both the compounds are elastically highly anisotropic. This should result in anisotropy in their mechanical behavior under applied stress.

**Table 6:** The maximum and minimum values of Young's modulus ($Y$ in GPa), compressibility ($\beta$ in TPa$^{-1}$), shear modulus ($G$ in GPa), Poisson's ratio ($\nu$) and their ratios for $B_6S$ and $B_6Se$ compounds.

| Compound | P (GPa) | Y | | B | | G | | ν ($10^{-3}$) | | $A_Y$ | $A_\beta$ | $A_G$ | $A_\nu$ | Ref. |
|---|---|---|---|---|---|---|---|---|---|---|---|---|---|---|
| | | $Y_{min}$ | $Y_{max}$ | $\beta_{min}$ | $\beta_{max}$ | $G_{min}$ | $G_{max}$ | $\nu_{min}$ | $\nu_{max}$ | | | | | |
| $B_6S$ | 0 | 229.07 | 453.82 | 1.65 | 3.28 | 88.53 | 203.21 | 9.35 | 31.27 | 1.98 | 1.99 | 2.29 | 33.44 | This work |
| | 5 | 240.75 | 490.07 | 1.42 | 2.91 | 90.96 | 215.96 | 20.67 | 337.70 | 2.04 | 2.05 | 2.37 | 16.33 | |
| | 10 | 217.10 | 513.57 | 1.22 | 2.76 | 77.19 | 222.61 | 9.83 | 420.96 | 2.37 | 2.25 | 2.88 | 42.84 | |
| | 15 | 268.3 | 542.97 | 1.15 | 2.39 | 99.77 | 233.21 | 41.15 | 356.72 | 2.02 | 2.09 | 2.34 | 8.67 | |
| | 20 | 291.11 | 568.07 | 1.05 | 2.14 | 108.40 | 242.33 | 63.51 | 354.26 | 1.95 | 2.04 | 2.23 | 5.58 | |
| $B_6Se$ | 5 | 294.87 | 464.26 | 1.48 | 2.91 | 134.63 | 197.59 | 28.18 | 172.65 | 1.57 | 1.96 | 1.47 | 6.13 | |
| | 10 | 367.50 | 506.11 | 1.31 | 2.08 | 164.74 | 208.46 | 108.75 | 175.29 | 1.38 | 1.58 | 1.26 | 1.61 | |
| | 15 | 374.49 | 532.09 | 1.20 | 2.05 | 169.24 | 218.11 | 98.45 | 187.57 | 1.42 | 1.71 | 1.29 | 1.91 | |
| | 20 | 397.57 | 551.67 | 1.11 | 1.83 | 171.33 | 225.97 | 120.08 | 197.42 | 1.39 | 1.66 | 1.32 | 1.64 | |

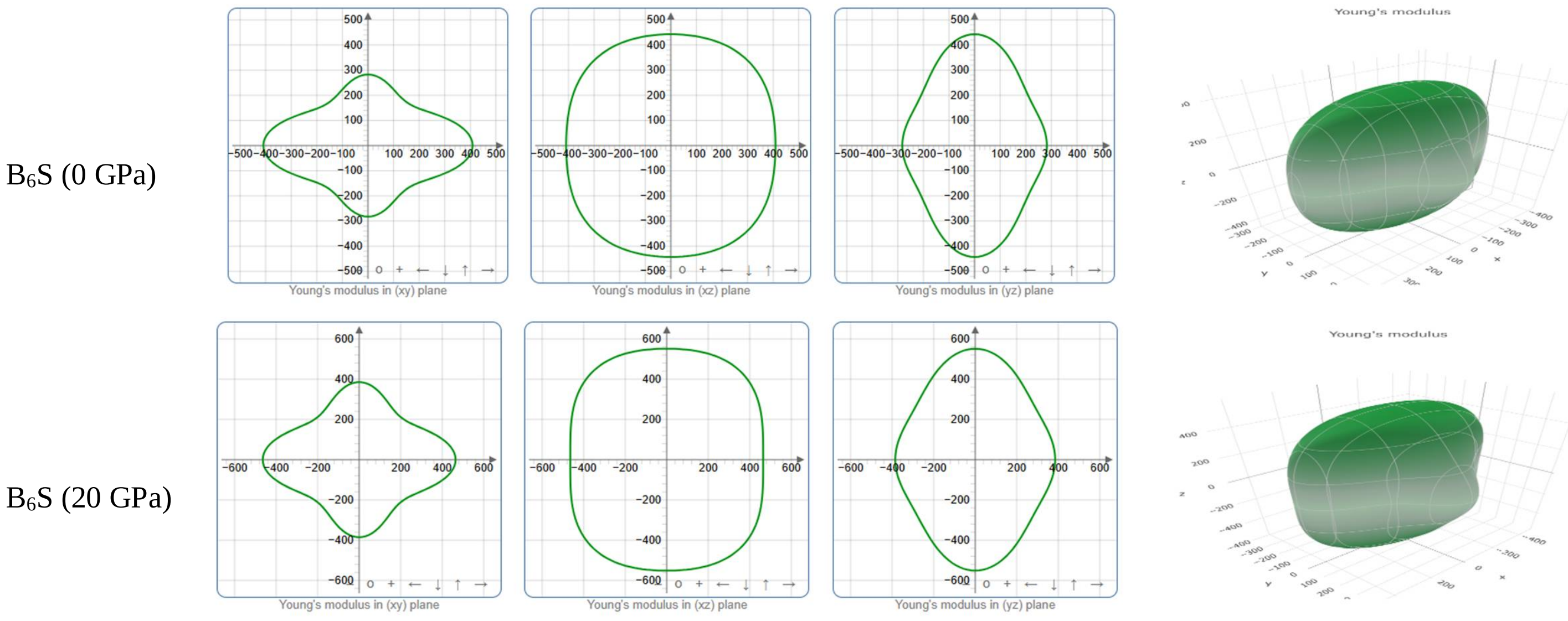

**Figure 7:** Directional variation in Young's modulus (Y) of $B_6S$ compound at 0 and 20 GPa.

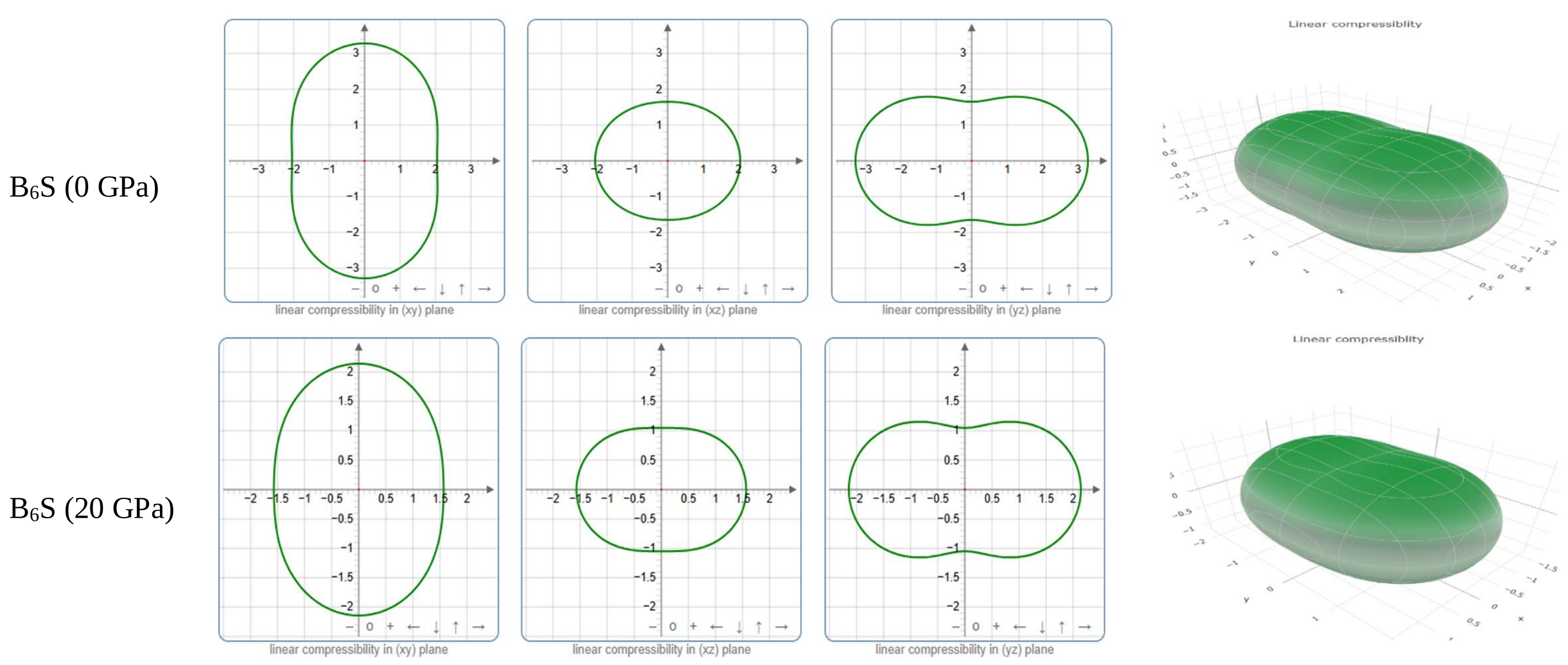

**Figure 8:** Directional variation in linear compressibility (β) of $B_6S$ compound at 0 and 20 GPa.

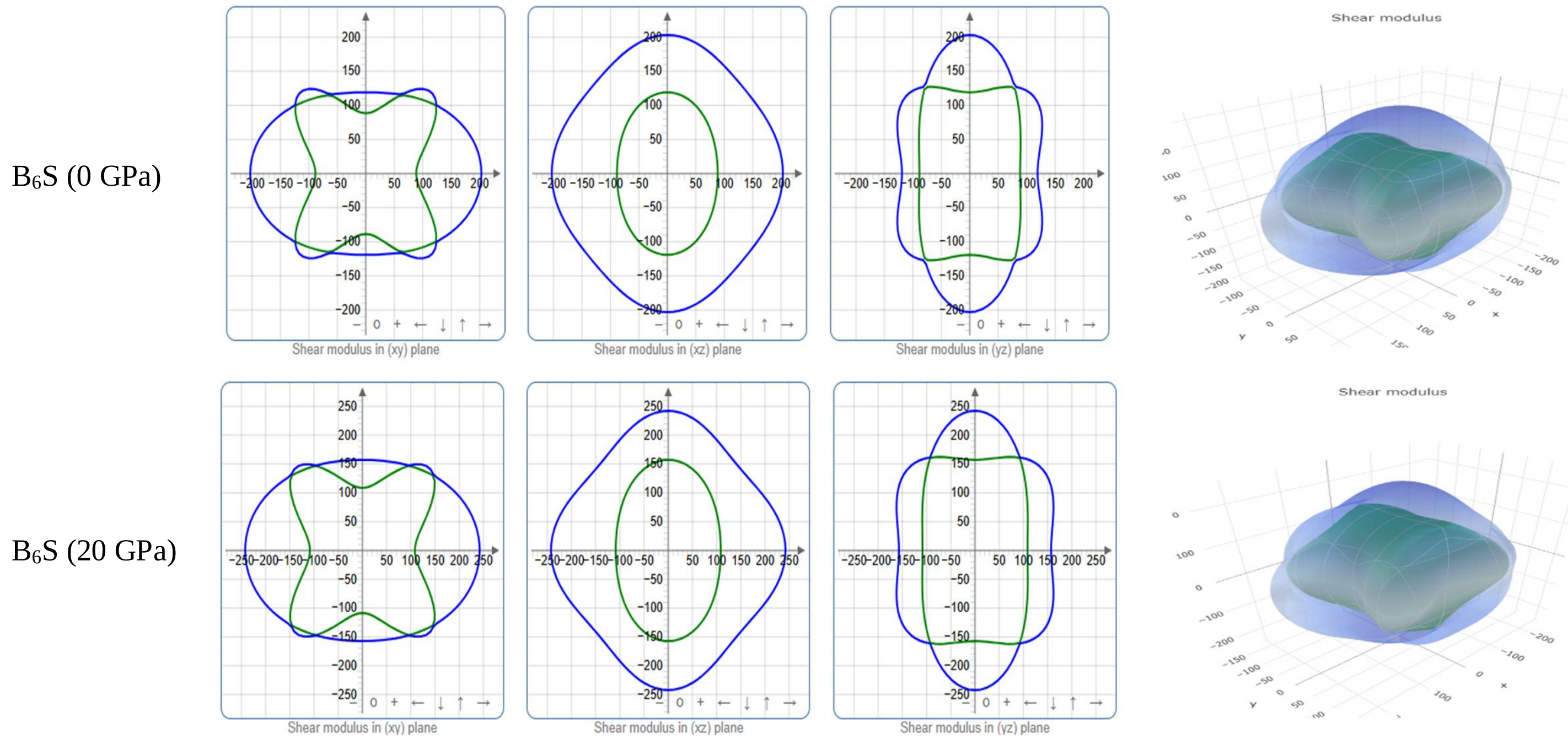

**Figure 9:** Directional variation in shear modulus (G) of $B_6S$ compound at 0 and 20 GPa.

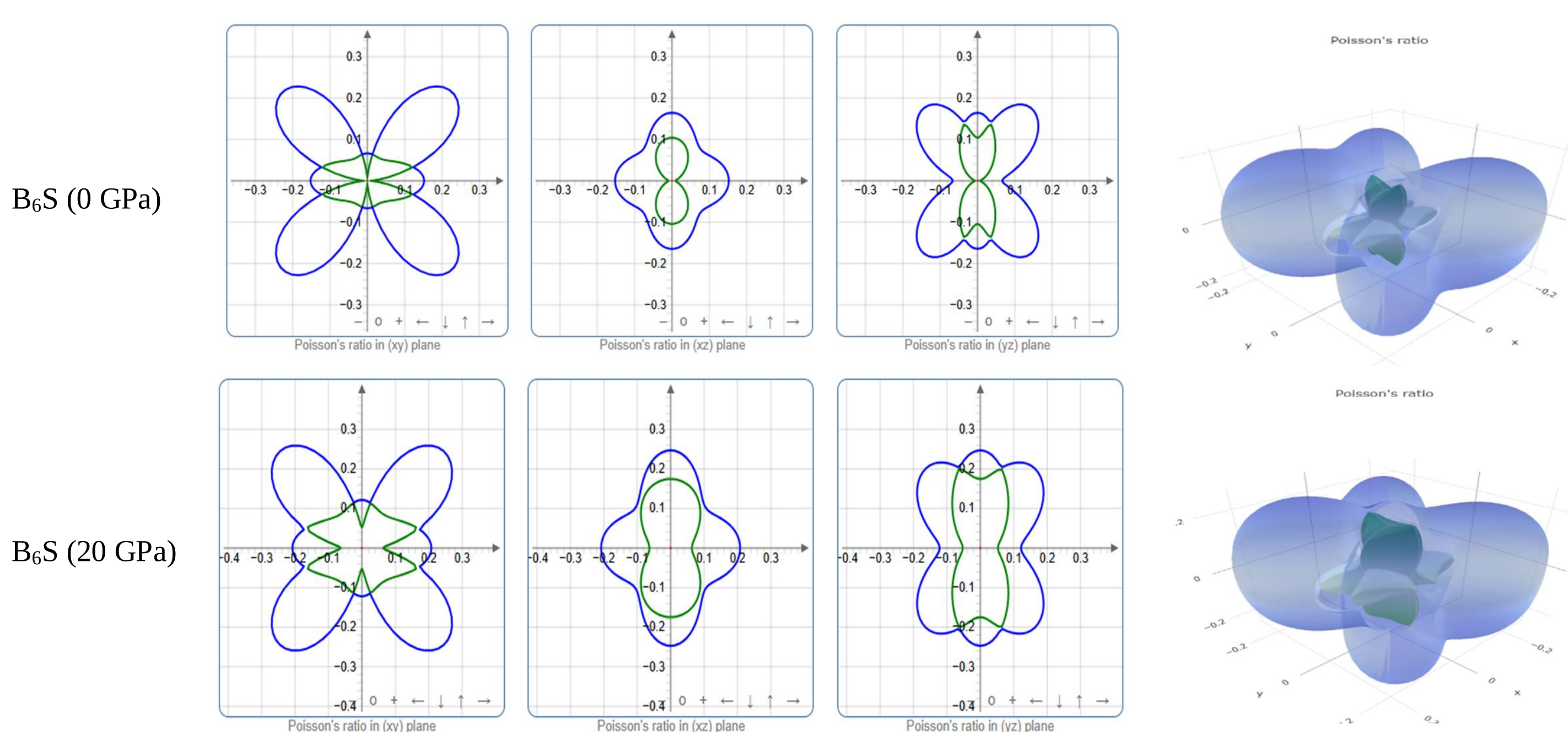

**Figure 10:** Directional variation in Poisson's ratio ($v$) of $B_6S$ compound at 0 and 20 GPa.

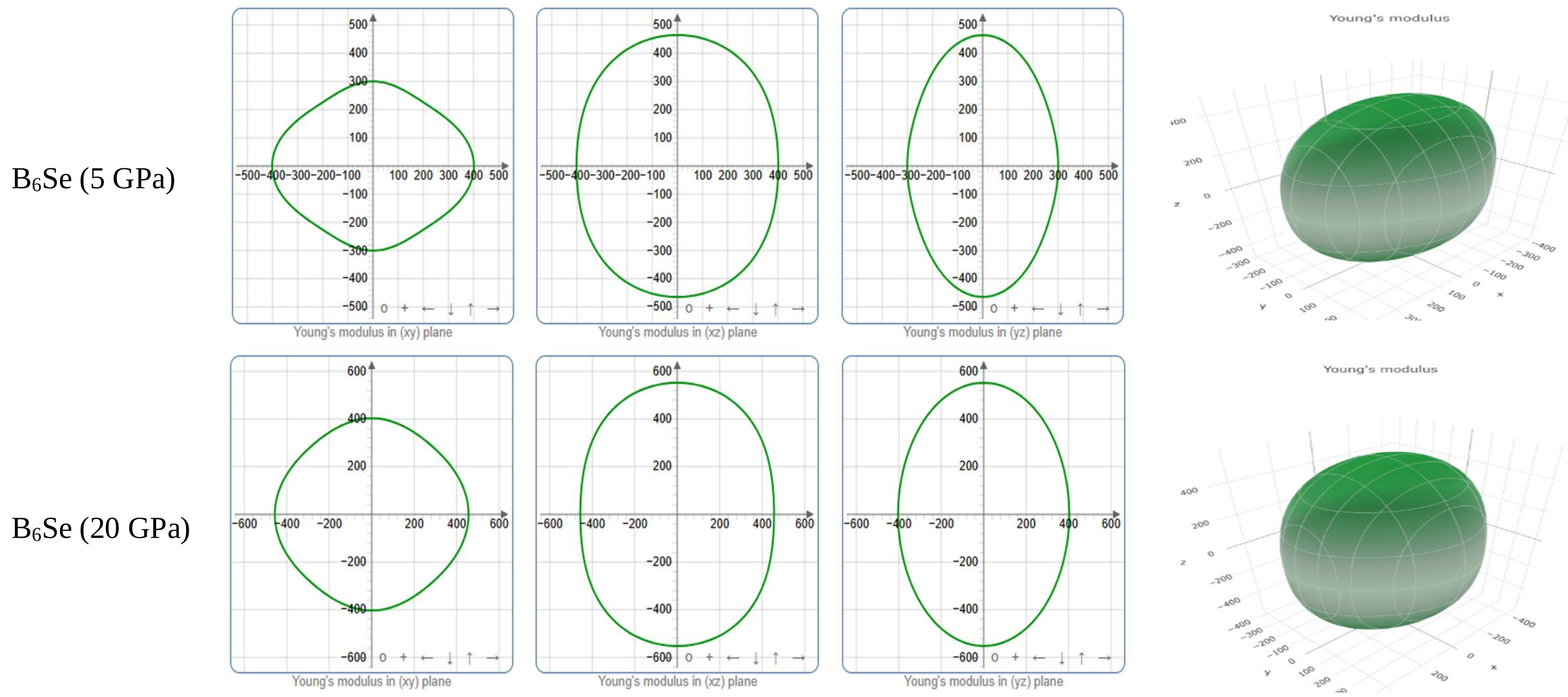

**Figure 11:** Directional variation in Young's modulus (Y) of $B_6Se$ compound at 5 and 20 GPa.

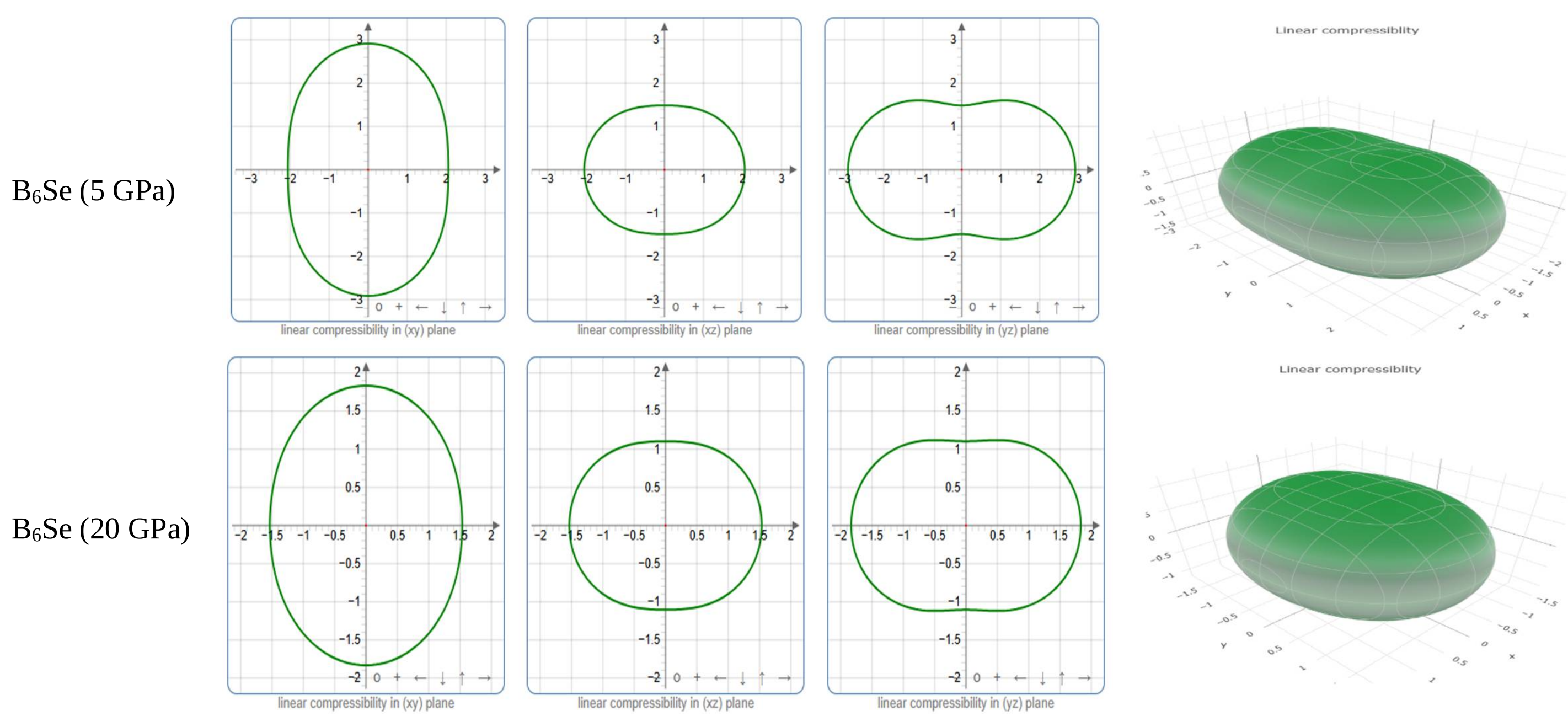

**Figure 12:** Directional variation in linear compressibility (β) of $B_6Se$ compound at 5 and 20 GPa.

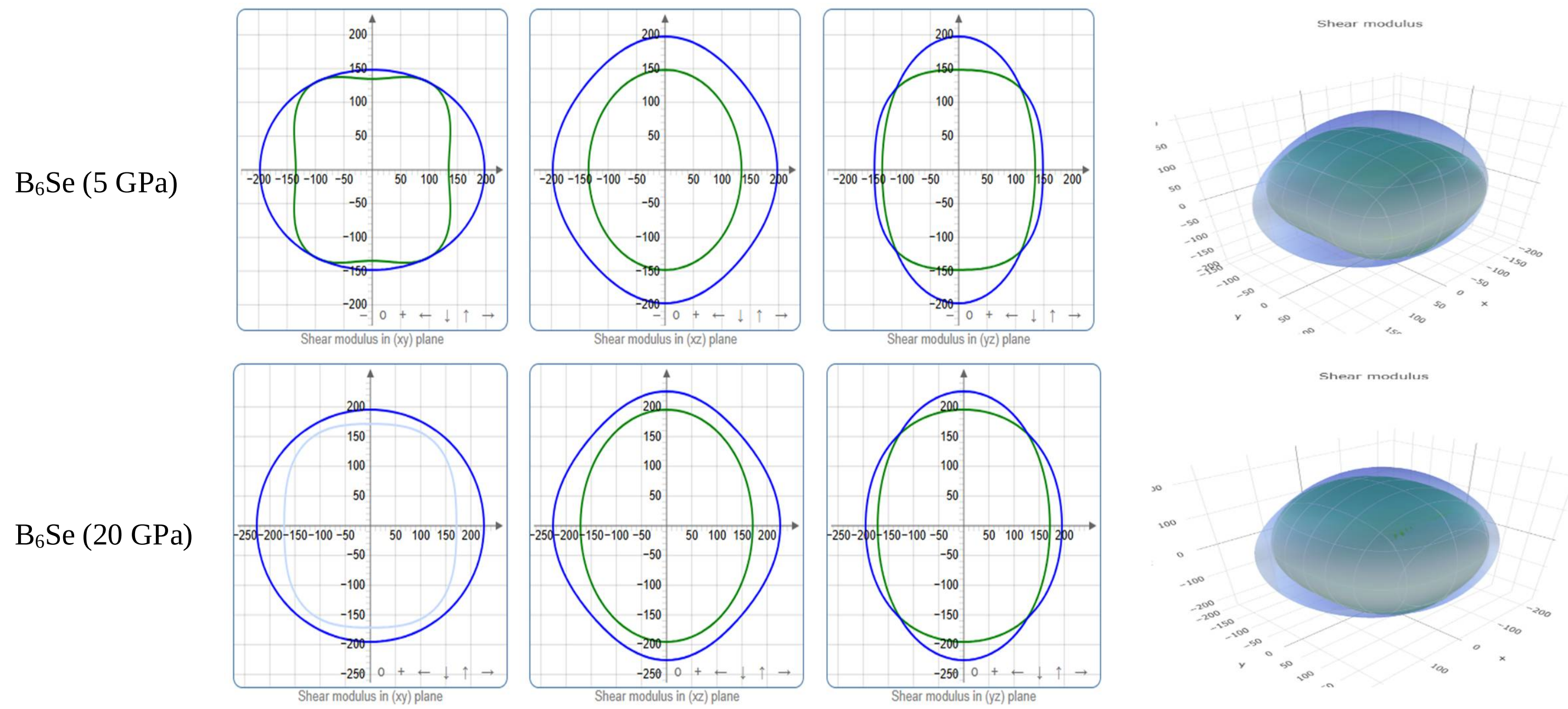

**Figure 13:** Directional variation in shear modulus (G) of $B_6Se$ compound at 5 and 20 GPa.

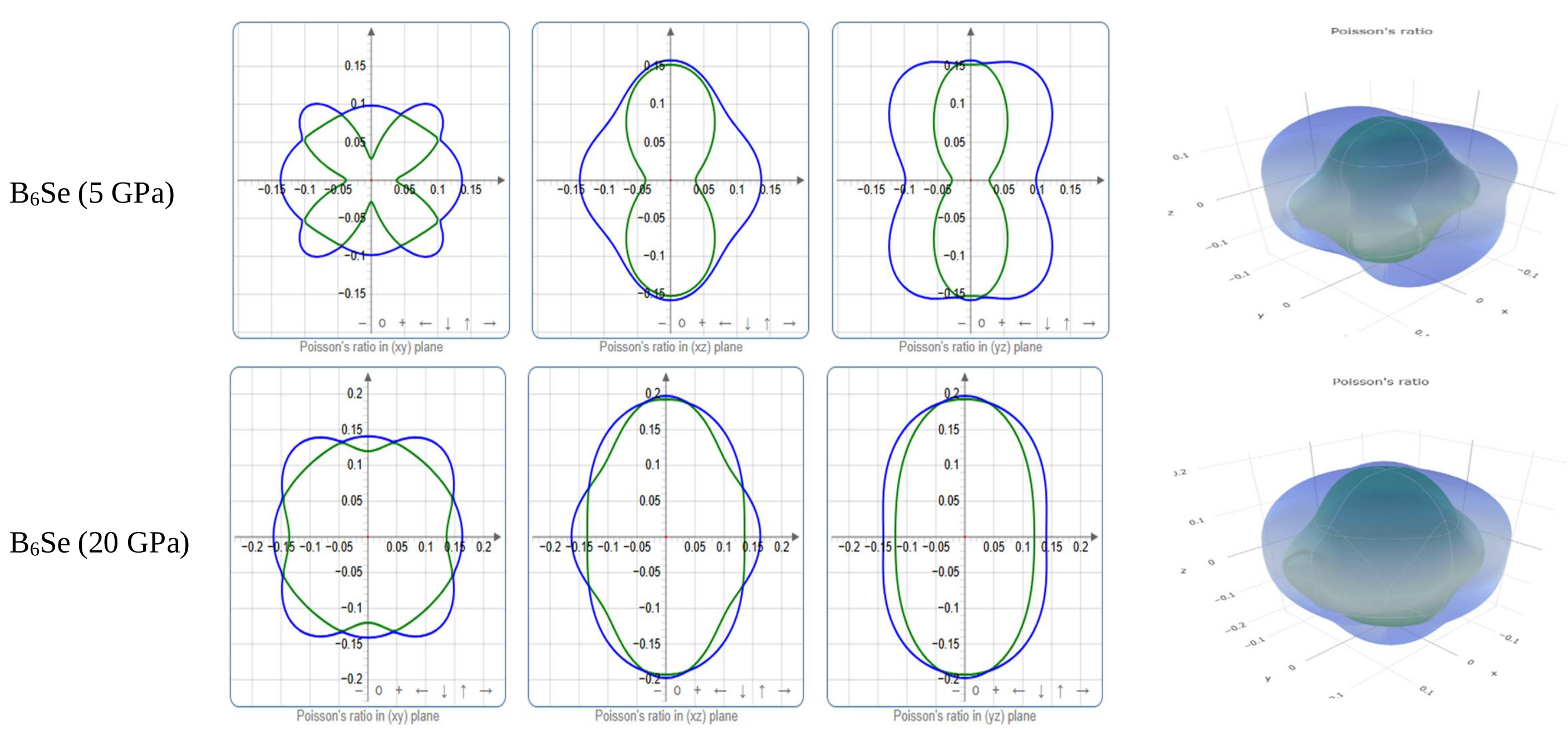

**Figure 14:** Directional variation in Poisson's ratio ($v$) of $B_6Se$ compound at 5 and 20 GPa.

### 3.3 Thermophysical Properties

#### (a) Debye and Melting Temperature

The Debye model effectively describes phonon contributions to the Gibbs energy of crystalline phases and uses lattice vibrations to explain the thermal behavior of solids. The Debye temperature divides the behavior at low and high temperatures as well as the classical and quantum phonon regimes. This one parameter can be used to describe a variety of lattice properties, including heat capacity, thermal expansion, thermal conductivity, compressibility, melting point, and lattice enthalpy. Every vibrational mode has the same energy, $k_B$T, at temperatures higher than $\theta_D$. For temperatures below $\theta_D$, however, higher frequency modes are regarded as frozen [71]. The predominant vibrational excitations in crystals at low temperatures are acoustic vibrations. The Debye temperature values obtained from specific heat measurements and elastic constants agree at low temperatures. We used an established method, the Anderson model, to calculate the Debye temperature of our compounds [72]:

$$\theta_D = \frac{h}{k_B}\left[\left(\frac{3n}{4\pi V_0}\right)\right]^{\frac{1}{3}} v_m \quad (13)$$

Here $h$ is Planck's constant, $k_B$ is Boltzmann's constant, $n$ is the number of atoms in the unit cell, $V_0$ is the equilibrium volume of the unit cell, and $v_m$ is the average sound velocity in the solid. The equation above shows that $\theta_D$ is proportional to the average sound velocity ($v_m$), which is determined by a crystal's elastic characteristics. The following equations can be used to determine $v_m$ based on the bulk ($B$) and shear ($G$) modulus and the longitudinal ($v_l$) and transverse ($v_t$) sound speeds [73]:

$$v_m = \left[\frac{1}{3}\left(\frac{2}{v_t^3} + \frac{1}{v_l^3}\right)\right]^{-\frac{1}{3}} \quad (14)$$

where,

$$v_t = \sqrt{\frac{G}{\rho}} \quad (15)$$

$$v_l = \sqrt{\frac{3B+4G}{3\rho}} \quad (16)$$

**Table 8** reports the calculated $\theta_D$ values of $B_6X$ (X = S, Se) at 0 K, showing that $B_6S$ possesses higher average sound velocity ($v_m$) and $\theta_D$ than $B_6Se$ because of its lower mass density. A lower $\theta_D$implies weaker phonon thermal conductivity; therefore, $B_6Se$ is expected to exhibit lower phonon thermal conductivity and is more suitable for thermal barrier coating applications. Additionally, $\theta_D$ increases with applied pressure (0-20 GPa), and for all cases, longitudinal sound velocities exceed transverse ones.

The melting temperature ($T_m$) is a key thermophysical parameter that defines a material's high-temperature applicability. A higher $T_m$ indicates greater cohesive energy and lower thermal expansion. Materials can safely operate below $T_m$ without oxidation, chemical degradation, or mechanical failure. In this work, the melting temperatures of $B_6X$ (X = S, Se) compounds are estimated from their elastic constants using the following relation [74]:

$$T_m = 354 + 1.5(2C_{11} + C_{33}) \tag{17}$$

**Table 7** presents the predicted melting temperatures ($T_m$) of the studied materials. Bonding strength is closely linked to lattice (phonon) thermal conductivity ($k_{ph}$) and $T_m$. $B_6Se$ exhibits a higher melting temperature due to its relatively larger elastic constants, indicating stronger bonding and cohesive energy than $B_6S$. These boron-rich chalcogenides, known as hard phase compounds, are therefore well suited for high-temperature applications. Materials with higher stiffness generally possess higher melting points, enabling them to withstand elevated operating temperatures, which mainly arise from a high heat of fusion, low entropy of fusion, or both.

**Table 7:** Calculated crystal density ($\rho$ in g/cm³), longitudinal, transverse and average sound velocities ($v_t$, $v_l$ and $v_m$ in km/s) and Debye temperature ($\theta_D$ in K) and melting temperature ($T_m$ in K) of $B_6X$ (X = S, Se) compounds for various pressure.

| Compound | P (GPa) | ρ | $v_l$ | $v_t$ | $v_m$ | $\theta_D$ | $T_m$ | Ref. |
|---|---|---|---|---|---|---|---|---|
| $B_6S$ | 0 | 15266.91 | 4728.06 | 3079.96 | 3376.77 | 1179.52 | 2299.80 | This work |
| | 5 | 15794.60 | 4835.72 | 3106.49 | 3411.52 | 1205.29 | 2425.35 | |
| | 10 | 16263.56 | 4826.39 | 3036.87 | 3342.95 | 1188.41 | 2507.59 | |
| | 15 | 16268.70 | 5086.85 | 3208.79 | 3531.21 | 1256.66 | 2634.87 | |
| | 20 | 17141.03 | 5127.83 | 3207.60 | 3533.16 | 1287.48 | 2732.16 | |
| $B_6Se$ | 5 | 22317.96 | 4117.48 | 2714.35 | 2971.55 | 1043.85 | 2311.70 | |
| | 10 | 23160.21 | 4403.27 | 2839.17 | 3116.59 | 1111.78 | 2531.87 | |
| | 15 | 23732.08 | 4422.88 | 2845.55 | 3124.40 | 1123.16 | 2616.65 | |
| | 20 | 24316.53 | 4493.41 | 2857.34 | 3141.59 | 1138.73 | 2694.42 | |

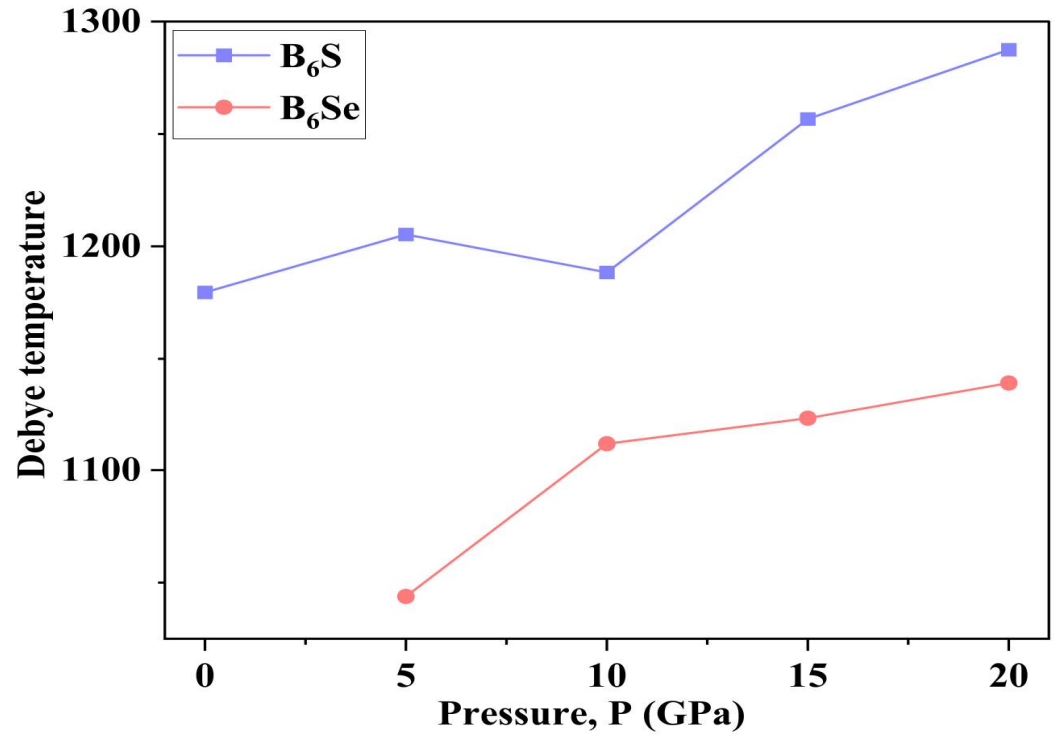

**Figure 15:** Debye temperature of $B_6S$ and $B_6Se$ for various pressures.

### (b) Lattice Thermal Conductivity

Lattice (phonon) thermal conductivity ($k_{ph}$) is an important property for applications such as thermoelectrics, heat sinks, and thermal barrier coatings [72,74]. It represents a material's ability to transfer heat through phonon transport in the crystal lattice. Materials with high $k_{ph}$are suitable for heat dissipation, whereas those with low $k_{ph}$ are effective thermal barriers. In this study, $k_{ph}$ for $B_6S$ and $B_6Se$ at 300 K was estimated using Slack's empirical model [75], with the results listed in **Table 8.**

$$k_{ph} = A\frac{M_{av}\Theta_D^3\delta}{\gamma^2 n^{2/3} T} \tag{18}$$

In the equation above, $M_{av}$ is the average atomic mass in kg/mol, $\theta_D$ is the Debye temperature in K, $\delta$ is the cubic root of average atomic volume in meter (m), $n$ is the number of atoms in the conventional unit cell, $T$ is the absolute temperature in K, and $\gamma$ is the acoustic Grüneisen parameter that measures the degree of anharmonicity of the phonons. A material with a low Grüneisen parameter value exhibits low phonon anharmonicity, resulting in strong thermal conductivity. The following equation may be used to determine Grüneisen parameter based on the Poisson's ratio [76]:

$$\gamma = \frac{3(1+\nu)}{2(2-3\nu)} \tag{19}$$

The factor A($\gamma$), due to Julian [77], is computed from:

$$A(\gamma) = \frac{5.720\times10^7\times0.849}{2\times(1-{0.514}/{\gamma}+{0.228}/{\gamma^2})} \tag{20}$$

**Table 8** shows the calculated lattice thermal conductivity at room temperature (300 K), as well as the Grüneisen parameter. A high value of the Grüneisen parameter leads to high crystal compressibility and coefficient of thermal expansion. Both $B_6S$ and $B_6Se$ have low Grüneisen parameter. $B_6S$ has a higher Debye temperature than $B_6Se$. But $B_6Se$ has a higher lattice thermal conductivity than $B_6S$. Higher thermal conductivity suggests stronger covalent bonds. Overall, the lattice thermal conductivity of these two compounds at 300 K are low. The estimated values indicate that covalent bonding in $B_6Se$ is substantially stronger than in the $B_6S$ compound.

### (c) Minimum Thermal Conductivity and its Anisotropy

The high-temperature behavior of solids above the Debye temperature is important for practical applications. In this regime, intrinsic thermal conductivity reaches a lower limit called the minimum thermal conductivity ($k_{min}$), which is independent of defects or impurities. Clarke derived an expression for estimating $k_{min}$at high temperatures using the quasi-harmonic Debye model [78]:

$$k_{min} = k_B \upsilon_m (V_{\text{atomic}})^{-\frac{2}{3}} \tag{21}$$

In this equation, $k_B$, $\upsilon_m$ and $V_{\text{atomic}}$ represent the Boltzmann constant, average sound velocity, and cell volume per atom, respectively. Materials with higher sound velocity, low phonon mean free path, and Debye temperature exhibit higher minimum thermal conductivity. **Table 8** shows the estimated isotropic minimum thermal conductivity values for $B_6S$ and $B_6Se$. To discuss the anisotropy of thermal conductivity, the lowest thermal conductivities of $B_6S$ and $B_6Se$ compounds along distinct crystal orientations are examined using the Cahill's model [79]:

$$k_{min} = \frac{k_B}{2.48} n^{2/3} (\upsilon_l + \upsilon_{t1} + \upsilon_{t2}) \tag{22}$$

where, $k_B$ is the Boltzmann constant, $n$ is the number of atoms per unit volume, and $N$ is the total number of atoms in the cell with volume $V$. The computed minimum thermal conductivity of both the compounds are extremely low.

### (d) Thermal Expansion Coefficient

Thermal expansion arises from anharmonic lattice vibrations and strongly influences properties such as thermal conductivity, specific heat, entropy, compressibility, and electronic behavior. Materials with low thermal expansion are therefore important both scientifically and technologically. In this work, the thermal expansion coefficient ($\alpha$) is calculated from the shear modulus using the following relation [34]:

$$\alpha = \frac{1.6 \times 10^{-3}}{G} \tag{23}$$

where, $G$ is the isothermal share modulus in GPa. The thermal expansion coefficient of a material is inversely linked to its melting temperature ($\alpha \approx 0.02/T_m$) [35]. Crystals' thermal expansion coefficients vary with temperature. **Table 8** shows the computed values for $B_6S$ and $B_6Se$ compounds at 300 K. Both $B_6Se$ and $B_6Se$ have exceptionally low thermal expansion coefficients. Low thermal expansion materials are widely used in high anti-thermal shock applications, electronic devices, heat-engine components, spintronic devices, insulating coating and so on.

**Table 8:** Lattice thermal conductivity ($k_{ph}$ in W/m. K) at 300 K, Grüneisen parameter ($\gamma$), thermal expansion coefficient ($\alpha$ in $K^{-1}$), and minimum thermal conductivity ($k_{min}$ in W/m. K) of $B_6X$ (X = S, Se) compounds for various pressure.

| Compound | P (GPa) | $K_{ph}$ | $K_{min}$ ($10^{-4}$) | $\gamma$ | $\alpha$ ($10^{-5}$) | Ref. |
|---|---|---|---|---|---|---|
| $B_6S$ | 0 | 1.30 | 2.86 | 1.06 | 1.10 | This work |
| | 5 | 1.24 | 2.96 | 1.11 | 1.05 | |
| | 10 | 1.02 | 2.96 | 1.19 | 1.07 | |
| | 15 | 1.23 | 3.13 | 1.18 | 0.96 | |
| | 20 | 1.22 | 3.24 | 1.21 | 0.91 | |
| $B_6Se$ | 5 | 1.47 | 2.49 | 1.01 | 0.97 | |
| | 10 | 1.49 | 2.68 | 1.09 | 0.86 | |
| | 15 | 1.50 | 2.73 | 1.10 | 0.83 | |
| | 20 | 1.43 | 2.80 | 1.15 | 0.81 | |

From the results summarized in **Tables 7 and 8**, it is seen that both the compounds have high potential to be employed as efficient thermal barrier coating (TBC) [80]. Moreover, TBC performance of the two compounds can be finely tuned via application of pressure [81].

### 3.4 Bond Population Analysis

Atomic charges and charge transfer in compounds are two often utilized concepts in condensed matter physics. Understanding bonding types (ionic, covalent, and metallic) in $B_6S$ and $B_6Se$ compounds, we have used both Mulliken population analysis (MPA) [82] and Hirshfeld population analysis (HPA) [83]. **Table 9** shows the findings of this investigation for the $B_6X$ (X = S, Se) compounds. Mulliken charge examines how the electronic structure changes in response to atomic displacements. This charge is also linked to the dipole moment, electric polarizability, electronic structure, charge mobility in chemical processes, and other features of crystal structures. The charge spilling parameter measures the number of valence charges absent from the orbital projection. The smaller the parameter value, the better the model of electronic bonding. For the predicted hard phase, $B_6S$ and $B_6Se$ compound, S and Se contain positive Mulliken charge and B contains negative Mulliken charge so electrons are transferred from the S and Se atoms to the B atom, which is consistent with the previously investigated $B_6S$ and $B_6Se$ compounds. The Mulliken charge value for S is +0.24e at 0 GPa and +0.28e at 20 GPa for $B_6S$ also for Se is +0.94e at 5 GPa and +1.01e at 20 GPa for $B_6Se$ compound. This suggests that an ionic contribution to the bonding is present in these compounds. Therefore, in $B_6S$ and $B_6Se$ compounds, we may state that the B-B, B-S and S-S bonds have a partly ionic character. The degree of covalency and/or ionicity may be deduced from the values of effective valence charge, which is defined as the difference between the formal ionic charge and the Mulliken charge on the cation species. It establishes whether a link is ionic or covalent in strength. A perfect ionic bond is formed when the effective valence charge is precisely equal to zero. Covalent bonding is indicated by an atom having an effective valence charge that is not zero. The degree of divergence from zero may be used to assess the covalent bonding level. In $B_6X$ (X = S, Se) compounds, the presence of both covalent and ionic bonds is implied by the non-zero values of EVC (see **Table 9**). The order of covalency level of the chemical bonds are: B-B > B-S > S-S in $B_6S$ and B-B > B-Se > Se-Se in $B_6Se$. The reverse of these orders should indicate the order of ionicity level of the chemical bonds in $B_6X$ (X = S, Se) compounds. Overall, the effect of pressure on the bonding characteristics is moderate within the range of pressure considered.

**Table 9:** Charge spilling parameter (%), orbital charges (electronic charge), atomic Mulliken charge (electronic charge), Formal ionic charge (electronic charge), Effective valence and Hirshfeld charge (electronic charge) of (a) 0 GPa and (b) 20 GPa for $B_6S$ and (c) 5 GPa and (d) 20 GPa for $B_6Se$ compounds.

**(a)**

| Compound | Charge spilling | Species | Mulliken atomic populations | | | Mulliken charge | Formal ionic charge | Effective valence | Hirshfeld charge | Effective valence |
|---|---|---|---|---|---|---|---|---|---|---|
| | | | s | p | Total | | | | | |
| $B_6S$ | 0.73 | B | 0.84 | 2.18 | 3.02 | -0.02 | -3 | -2.98 | -0.04 | -2.96 |
| | | | 0.80 | 2.26 | 3.06 | -0.06 | | -2.94 | -0.02 | -2.98 |
| | | | 0.83 | 2.20 | 3.03 | -0.03 | | -2.97 | -0.05 | -2.95 |
| | | | 0.79 | 2.26 | 3.05 | -0.05 | | -2.95 | -0.04 | -2.96 |
| | | S | 1.72 | 4.05 | 5.76 | 0.24 | 1 | 0.76 | 0.21 | 0.79 |

**(b)**

| Compound | Charge spilling | Species | Mulliken atomic populations | | | Mulliken charge | Formal ionic charge | Effective valence | Hirshfeld charge | Effective valence |
|---|---|---|---|---|---|---|---|---|---|---|
| | | | s | p | Total | | | | | |
| $B_6S$ | 0.71 | B | 0.80 | 2.23 | 3.03 | -0.03 | -3 | -2.97 | -0.04 | -2.96 |
| | | | 0.77 | 2.29 | 3.06 | -0.06 | | -2.94 | -0.02 | -2.98 |
| | | | 0.76 | 2.29 | 3.05 | -0.04 | | -2.96 | -0.07 | -2.93 |
| | | S | 1.71 | 4.02 | 5.73 | 0.28 | 1 | 0.72 | 0.23 | 0.77 |

**(c)**

| Compound | Charge spilling | Species | Mulliken atomic populations | | | Mulliken charge | Formal ionic charge | Effective valence | Hirshfeld charge | Effective valence |
|---|---|---|---|---|---|---|---|---|---|---|
| | | | s | p | Total | | | | | |
| $B_6Se$ | 0.53 | B | 0.87 | 2.24 | 3.11 | -0.11 | -3 | -2.89 | -0.04 | -2.96 |
| | | | 0.85 | 2.34 | 3.20 | -0.20 | | -2.80 | -0.04 | -2.96 |
| | | | 0.86 | 2.27 | 3.13 | -0.13 | | -2.87 | -0.05 | -2.95 |
| | | | 0.84 | 2.35 | 3.19 | -0.19 | | -2.81 | -0.06 | -2.94 |
| | | Se | 1.13 | 3.93 | 5.06 | 0.94 | 1 | 0.06 | 0.26 | 0.74 |

**(d)**

| Compound | Charge spilling | Species | Mulliken atomic populations | | | Mulliken charge | Formal ionic charge | Effective valence | Hirshfeld charge | Effective valence |
|---|---|---|---|---|---|---|---|---|---|---|
| | | | s | p | Total | | | | | |
| $B_6Se$ | 0.51 | B | 0.85 | 2.28 | 3.13 | -0.13 | -3 | -2.87 | -0.04 | -2.96 |
| | | | 0.84 | 2.37 | 3.21 | -0.20 | | -2.80 | -0.04 | -2.96 |
| | | | 0.82 | 2.33 | 3.15 | -0.15 | | -2.85 | -0.07 | -2.93 |
| | | | 0.82 | 2.36 | 3.18 | -0.18 | | -2.82 | -0.06 | -2.94 |
| | | Se | 1.08 | 3.91 | 4.99 | 1.01 | 1 | -0.01 | 0.28 | 0.72 |

### 3.5 Charge Density Distribution

The electronic charge density distribution map displays precise information about the bonding type and amount of charge transfer between atoms in a molecule. It depicts the formation and depletion of electronic charges adjacent to various atomic species. The collection of charges between two atoms demonstrates covalent bonding between them. The presence of ionic bonds may be anticipated using a negative and positive charge balance at the atom locations. To get a better understanding of chemical bonding between the ions of $B_6S$ and $B_6Se$ compounds, we examined the electronic charge density distributions depicted in **Figures. 16 and 17** in the crystallographic plane (100). The color scales show the intensity of total electronic density (e/Å$^3$). The colors blue and red signify high and low electronic densities, whereas red and blue represent high and low electronic densities in $B_6S$ and $B_6Se$ compounds, respectively. **Fig. 16** shows that S has the largest charge density and B has the lowest charge density. **Fig. 17** shows, Se has the highest charge density, and B has the lowest charge density. Thus, the charge density distributions in $B_6S$ and $B_6Se$ compounds are anisotropic. Furthermore, the uneven charge distributions around the atoms are not circular. There are directional dependencies. This means that atomic bonding has both ionic and covalent components. The overall charge distribution in $B_6S$ and $B_6Se$ is qualitatively similar to that of $B_6S$ and $B_6Se$ [28,29]. It is evident that the blue color of the compounds indicates a significant reduction in electron density around the S/Se atomic species, whereas the red color indicates a build-up of charge around the B element. This implies that the S/Se atoms transfer their charge to the B atoms. This shows that there is an ionic connection. The B-B and B-S/Se atoms simultaneously exhibit a substantial and directed accumulation of charge, indicating the development of strong covalent bonds.

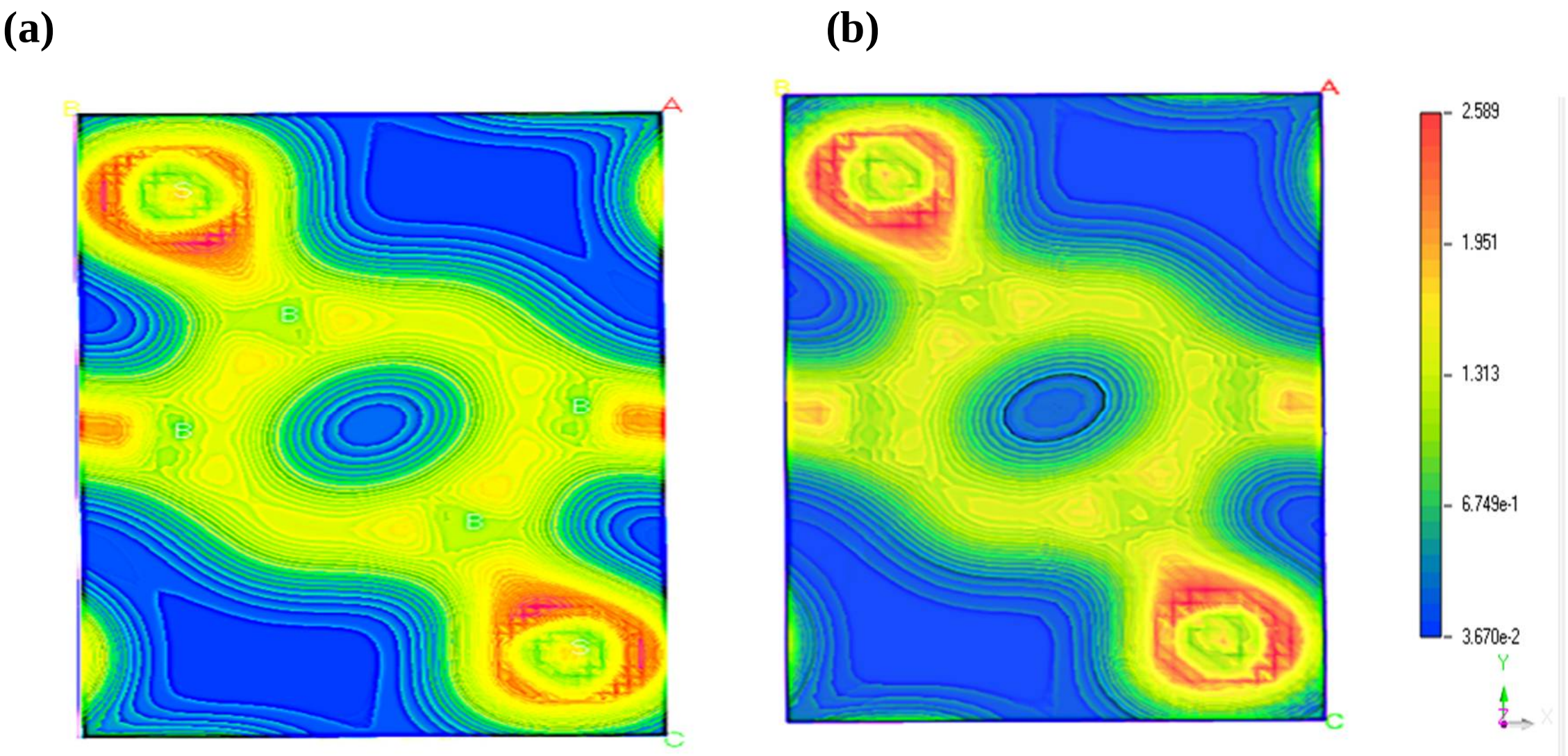

**Figure 16:** The electronic charge density map in the (100) plane of $B_6S$ compound at (a) 0 GPa and (b) 20 GPa. The charge density scale is shown on the right.

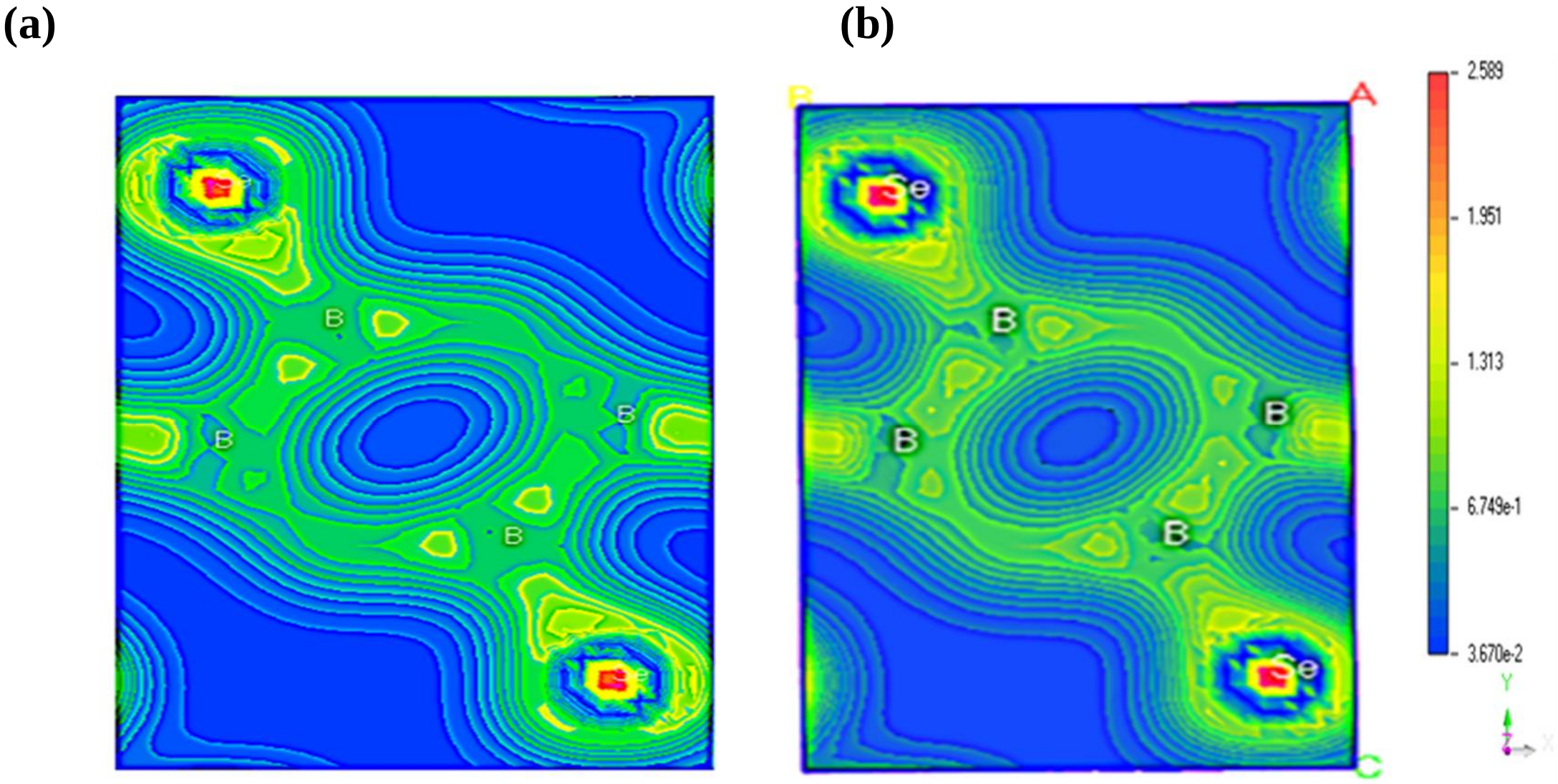

**Figure 17:** The electronic charge density map in the (100) plane of $B_6Se$ compound at (a) 5 GPa and (b) 20 GPa. The charge density scale is shown on the right

### 3.6 Optical Properties

The material's optical properties were investigated to understand how it interacts with electromagnetic radiation and its potential for optoelectronic applications. All-important photon energy-dependent optical features are governed by the underlying electronic band structure of the system of interest. The presence of a medium in electric and magnetic fields can cause induced current, polarization charges, electric dipoles, and magnetic moments [84]. Since both $B_6Se$ and $B_6S$ are elastically anisotropic, optical parameter may also show direction dependence. So, we have investigated optical properties of $B_6Se$ and $B_6S$ for three different polarization directions of the electric field [100], [010] and [001] under varying hydrostatic pressure. The energy-dependent optical properties examined include dielectric function $\varepsilon(\omega)$, refractive index $n(\omega)$, optical conductivity $\sigma(\omega)$, reflectivity $R(\omega)$, absorption coefficient $\alpha(\omega)$, and energy loss function $L(\omega)$ (where $\omega = 2\pi f$ is the angular frequency) for incident photon energies up to 30 eV.

The absorption coefficient $\alpha(\omega)$ reveals how much light of a particular energy (wavelength/frequency) may pass through a material before being absorbed and offers information on the best solar energy conversion efficiency. The absorption spectra of $B_6S$ and $B_6Se$ are shown in **Figures. 18** and **19**, respectively, reveal the insulating nature of the compounds since the spectra does not start from 0 eV. The absorption spectra of $B_6S$ and $B_6Se$ exhibit strong anisotropy and a pronounced response in the ultraviolet region under applied pressure. Along the [100] direction, both compounds show absorption maxima around 10-10.5 eV, with a slight blue shift as pressure increases, indicating pressure-induced modification of inter-band transitions. In the [010] direction, the absorption peaks move more significantly toward higher energies, reaching up to ~12.6 eV for $B_6S$ and ~11.9 eV for $B_6Se$ at 20 GPa, suggesting enhanced pressure sensitivity. A similar UV-dominated behavior is observed for the [001] direction, where the peak positions remain within 12.0-12.3 eV for $B_6S$ and 11.5-11.9 eV for $B_6Se$. These absorption features originate from direct electronic transitions between valence and conduction bands, primarily governed by pressure-induced band broadening and orbital hybridization. Since all prominent peaks lie well within the ultraviolet energy range (3.1-12.4 eV), both $B_6S$ and $B_6Se$ emerge as promising candidates for pressure-tunable UV optoelectronic applications. The absorption coefficient drops down quickly at the plasma edge, at around 20 eV for both the compounds. The absorption coefficient exhibits a moderate level of optical anisotropy.

The conduction of free charge carriers over a specific range of photon energy can be explained by optical conductivity $\sigma(\omega)$. Specifically, the peak intensities rise as their positions move toward higher photon energies, indicating increased electronic overlap and band structure changes brought on by pressure as **Figures. 18** and **19** illustrate.

The dielectric function $\varepsilon(\omega)$ shows how a material interacts with the incident electromagnetic waves. The macroscopic electronic response of a material can fully be described by the complex dielectric function:

$$\varepsilon(\omega) = \varepsilon_1(\omega) + i\varepsilon_2(\omega) \tag{24}$$

where $\varepsilon_1(\omega)$ and $\varepsilon_2(\omega)$ are real and imaginary parts of dielectric function $\varepsilon(\omega)$. In the condensed matter system, there are two contributions to $\varepsilon(\omega)$, namely intra-band and inter-band transitions. Intra-band transitions play an important role at low energy, and the inter-band term depends strongly upon the details of the electronic band structure [85]. For semiconductors in the ground state, intraband transitions are frozen. For $B_6S$, the principal $\varepsilon_1(\omega)$ peaks appear in the range of ~4.7-5.1 eV, while $\varepsilon_2(\omega)$ peaks lie between ~5.7 and 6.5 eV along the [001], [010], and [100] directions of electric field polarization, showing minor pressure-induced shifts from 0 to 20 GPa. At higher photon energies, both $\varepsilon_1$ and $\varepsilon_2$ decrease monotonically. In $B_6Se$, $\varepsilon_1(\omega)$ peaks occur at lower energies (~4.2-5.1 eV), whereas $\varepsilon_2(\omega)$ peaks are located around ~5.5-6.3 eV, with small but noticeable pressure-driven shifts between 5 and 20 GPa, reflecting changes in the electronic band structure. For the imaginary part $\epsilon_2(\omega)$, the peaks appear around 6.64 and 6.51 eV at 5 and 20 GPa respectively. For both compounds, the imaginary portion approaches zero at about ~27 eV. It implies that in this region these materials become almost transparent with very little absorption. At this point, the reflectivity spectra started to decrease, as shown in **Figures 18** and **19**. Within the pressure range investigated, this tendency is seen to continue.

The loss function $L(\omega)$ is a parameter that describes the energy loss of a fast electron via exciting collective charge oscillation modes while traversing a material [86,87]. The loss function $L(\omega)$, for $B_6S$ and $B_6Se$ phases is displayed in **Figures. 18** and **19** respectively. The energy-loss spectra of $B_6S$ and $B_6Se$ display pronounced, pressure-dependent plasmon peaks with strong directional variation. For $B_6S$, the peak energies increase from 24.42 to 25.71 eV for [001], from 24.16 to 25.25 eV for [010], and from 23.70 to 25.25 eV for [100] directions as pressure rises from 0 to 20 GPa. In comparison, $B_6Se$ shows corresponding shifts from 25.34 to 25.91 eV ([001]), 24.64 to 26.00 eV ([010]), and 25.34 to 26.70 eV ([100]) over 5 to 20 GPa pressure range. The distinct pressure-induced shifts along different directions clearly demonstrate the optical anisotropy of both compounds. The material's plasma frequency ($\omega_P$) is the frequency that corresponds to the peak of the loss spectrum, which represents the plasma resonance caused by collective charge excitation. At $\varepsilon_2 < 1$ and $\varepsilon_1 = 0$, this manifests [88,89]. The material becomes transparent to incident light when the incident photon frequency exceeds $\omega_P$.

**Figures. 18** and **19** respectively display the reflectivity spectra of $B_6S$ and $B_6Se$ as a function of photon energy. The reflectivity spectra of $B_6S$ and $B_6Se$ exhibit pronounced peaks for the [001], [010], and [100] directions under applied pressure. $B_6S$ shows higher-energy reflectivity peaks compared to $B_6Se$, with all peaks shifting toward higher energies as pressure increases which show strong directional and pressure dependences. For $B_6S$, the maximum reflectivity peak appears along the [100] direction at 20 GPa at an energy of 22.94 eV. In contrast, $B_6Se$ exhibits its highest reflectivity peak along the [010] direction at 20 GPa, at 15.60 eV. Overall, the reflectivity remains high in the ultraviolet region and non-selective in the ultraviolet energy range (10-23 eV). Optical anisotropy is small in the reflectivity spectra.

The refractive index is a complex parameter, expressed as $N(\omega) = n(\omega) + ik(\omega)$, where the imaginary part $k(\omega)$ is known as extinction coefficient. The imaginary part indicates the attenuation of electromagnetic wave when it passes through the media. The real part of the refractive index illustrates the phase velocity of the electromagnetic wave inside the sample. The real and imaginary parts of the refractive index as a function of photon energy are illustrated in **Figures. 18** and **19**. The $n(0)$ (static) values of $B_6S$ materials are higher for [001] and [100] directions compared to [010] directions at 0 GPa, at the same time in 10 and 20 GPa the [001] direction is higher as compared to [100] and [010] directions. Again, the $n$ (0) (static) values of $B_6Se$ materials are higher for [001] direction compared to [010] and [100] directions at 5, 10, and 20 GPa pressure. It can be concluded from the plots that $n(\omega)$ decreases sharply in the energy region 5.10 to ~15 eV. On the other hand, the extinction coefficient, $k(\omega)$ after reaching its maxima in the UV region, decreases gradually at higher energies. The effect of pressure on the complex refractive index is weak.

**(a)**

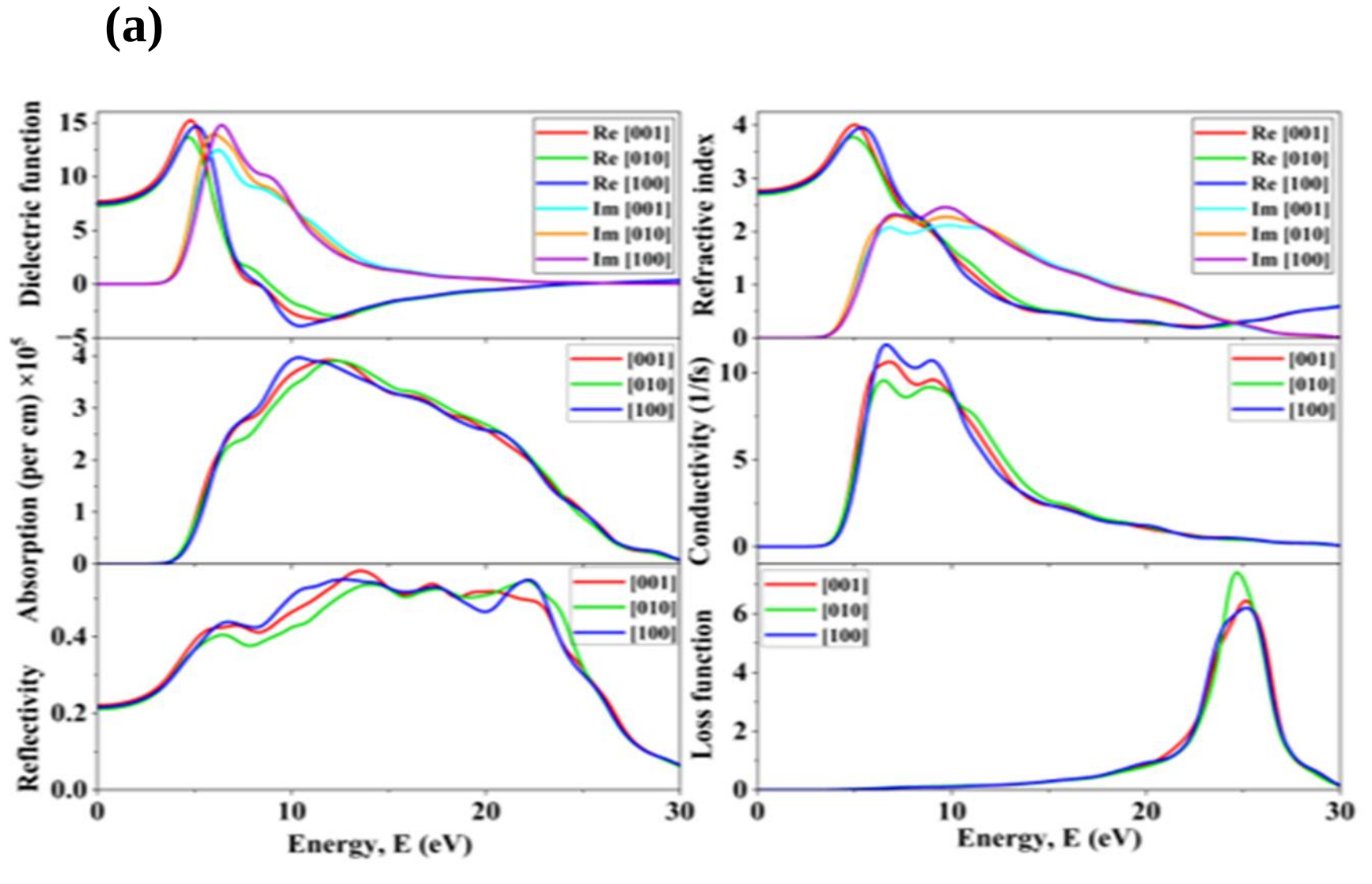
Dielectric function
Re [001]
Re [010]
Re [100]
Im [001]
Im [010]
Im [100]
Refractive index
Absorption (per cm) ×10^5
[001]
[010]
[100]
Conductivity (1/fs)
Reflectivity
Loss function
Energy, E (eV)

**(b)**

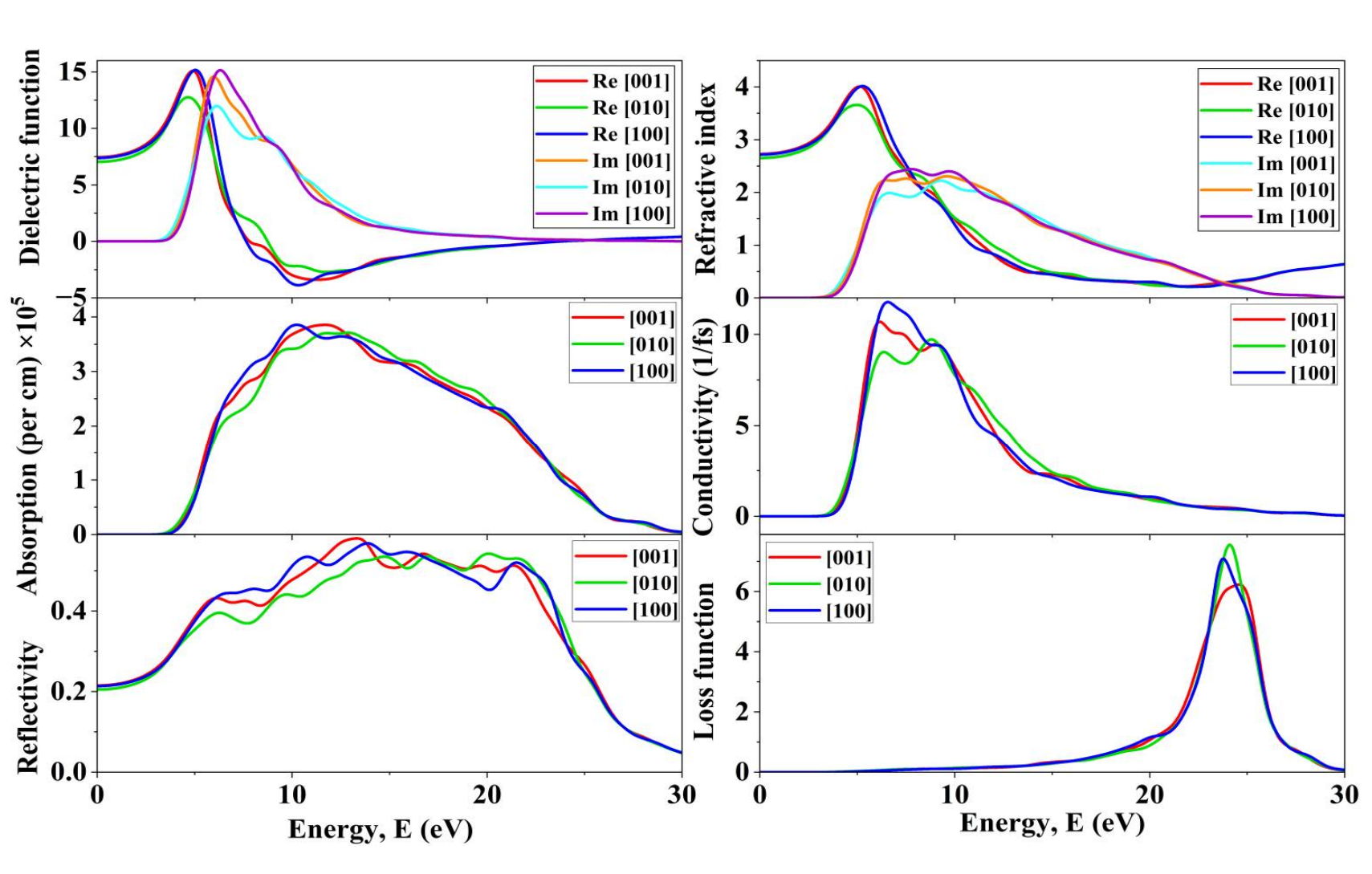
Dielectric function
Re [001]
Re [010]
Re [100]
Im [001]
Im [010]
Im [100]
Refractive index
Absorption (per cm) ×10^5
[001]
[010]
[100]
Conductivity (1/fs)
Reflectivity
Loss function
Energy, E (eV)

**(c)**

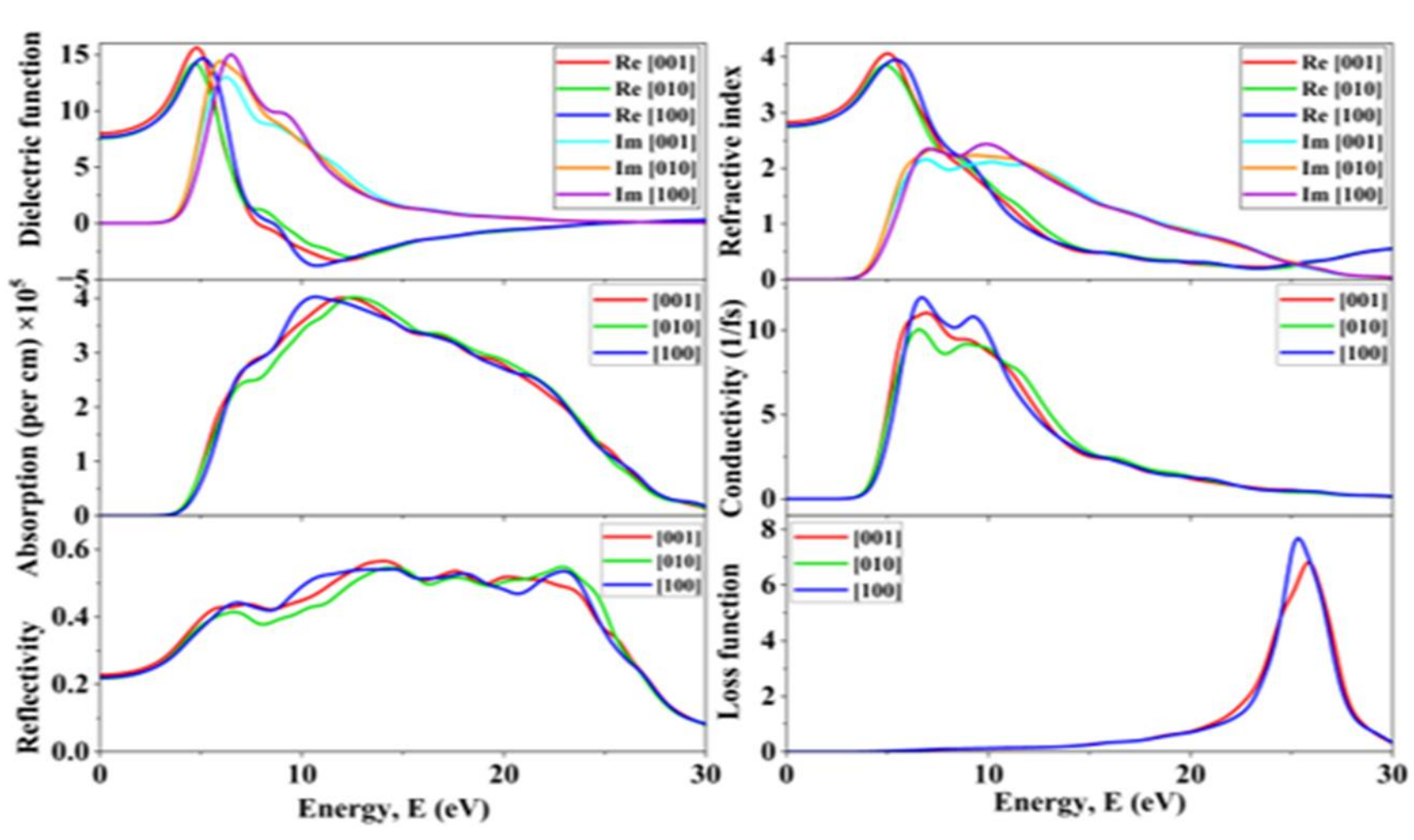
Dielectric function
Re [001]
Re [010]
Re [100]
Im [001]
Im [010]
Im [100]
Refractive index
Absorption (per cm) ×10^5
[001]
[010]
[100]
Conductivity (1/fs)
Reflectivity
Loss function
Energy, E (eV)

**Figure 18:** The energy dependent dielectric function, absorption coefficient, reflectivity, refractive index, optical conductivity, and loss function of $B_6S$ with electric field polarization vectors along [100], [010] and [001] directions at (a) 0 GPa (b) 10 GPa and (c) 20 GPa pressure.

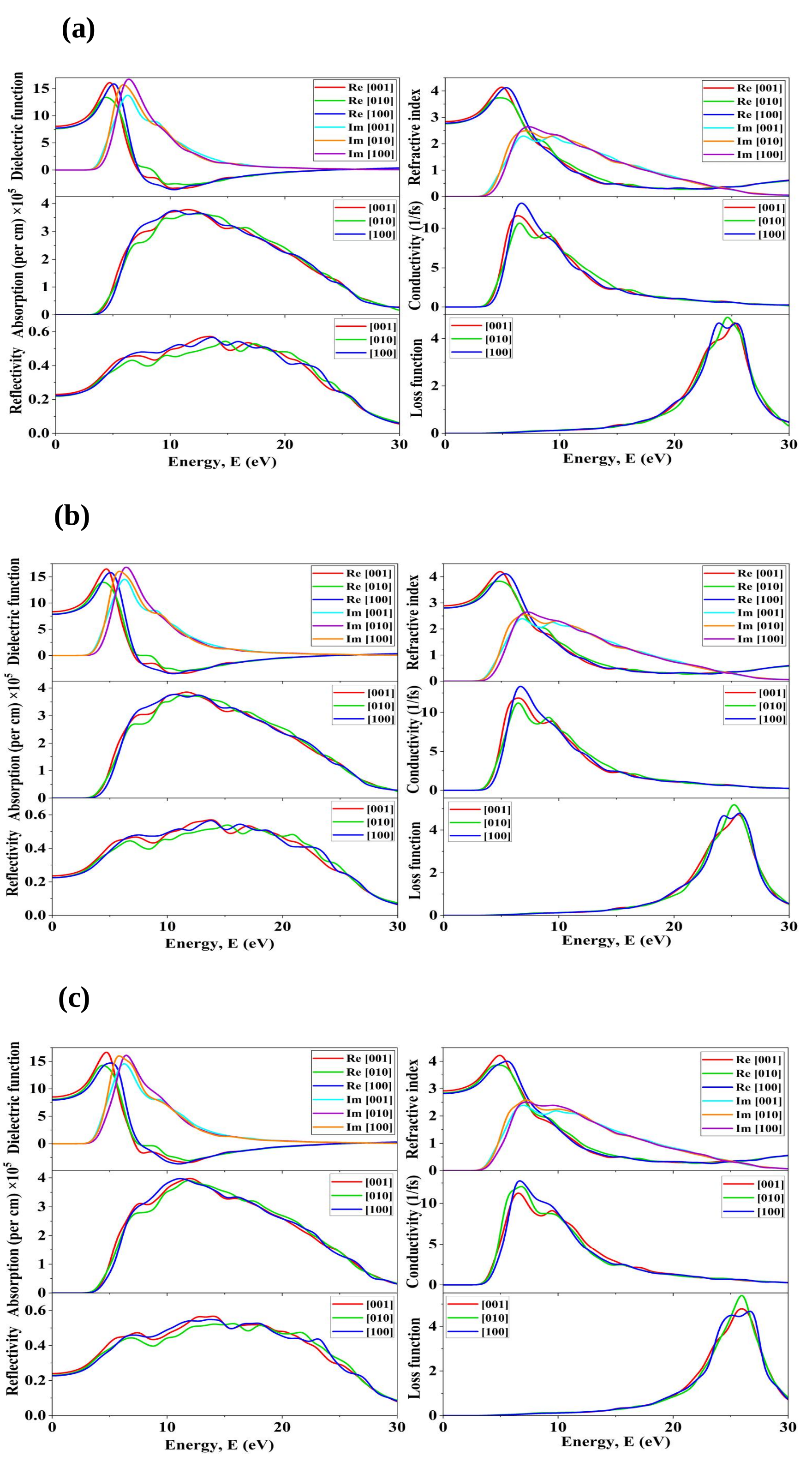

**Figure 19:** The energy dielectric function, absorption coefficient, reflectivity, refractive index, optical conductivity, and loss function of $B_6Se$ with electric field polarization vectors along [100], [010] and [001] directions at (a) 5 GPa (b) 10 GPa and (c) 20 GPa pressure.

### 3.7 Electronic properties

#### (a) Electronic band structure

At a microscopic level, the study of a material's electronic band structure is most important for explaining many of its features, such as chemical bonding, electronic transport, superconductivity, optical response, and magnetic order [90]. The band structure exhibits how the energy ($E$) of the allowed electronic states changes with the momentum ($k$) in the reciprocal lattice space. These $E(k)$ plots within the Brillouin zone (BZ) are also known as the electronic dispersion curves [91]. The electronic band structure for $B_6S$ and $B_6Se$ compounds for 0, 10, 20 GPa and 5, 10, 20 GPa, respectively, along high symmetry directions (G - Z - T - Y - S - X - U - R) of the first BZ in the energy range of −5 to +10 eV are shown in **Figures 20** and **21**. The Fermi level $E_F$ is illustrated as a horizontal dotted line in **Figures 20** and **21**. There is no overlap of conduction and valence bands and both the compounds show wide bandgap semiconducting character. Electronic band structure calculations reveal that $B_6S$ and $B_6Se$ are wide indirect bandgap semiconductors, with the valence band maximum and conduction band minimum occur at different $k$-points in the Brillouin zone [95].

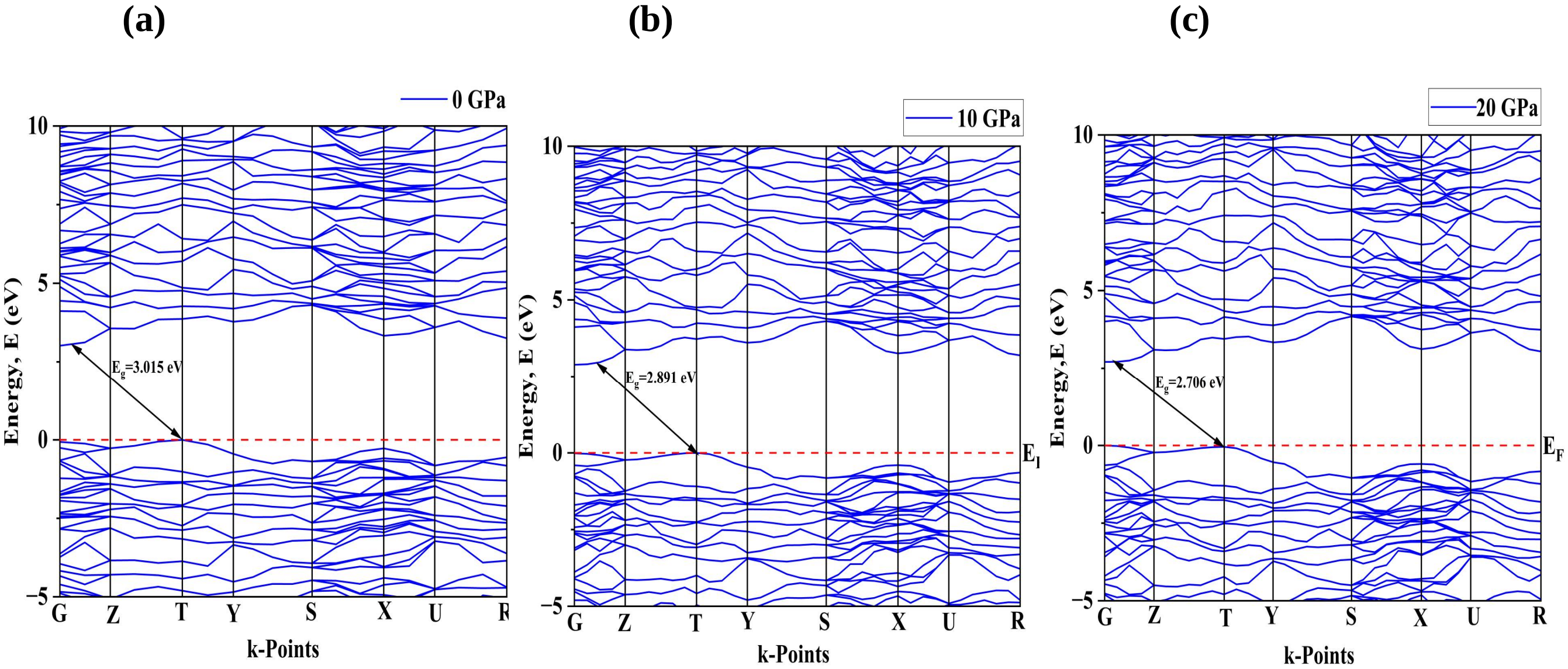

**Figure 20:** The calculated electronic band structures of the $B_6S$ compound (a) 0 GPa, (b) 10 GPa and (c) 20 GPa along several high symmetry directions of the Brillouin zone in the ground state.

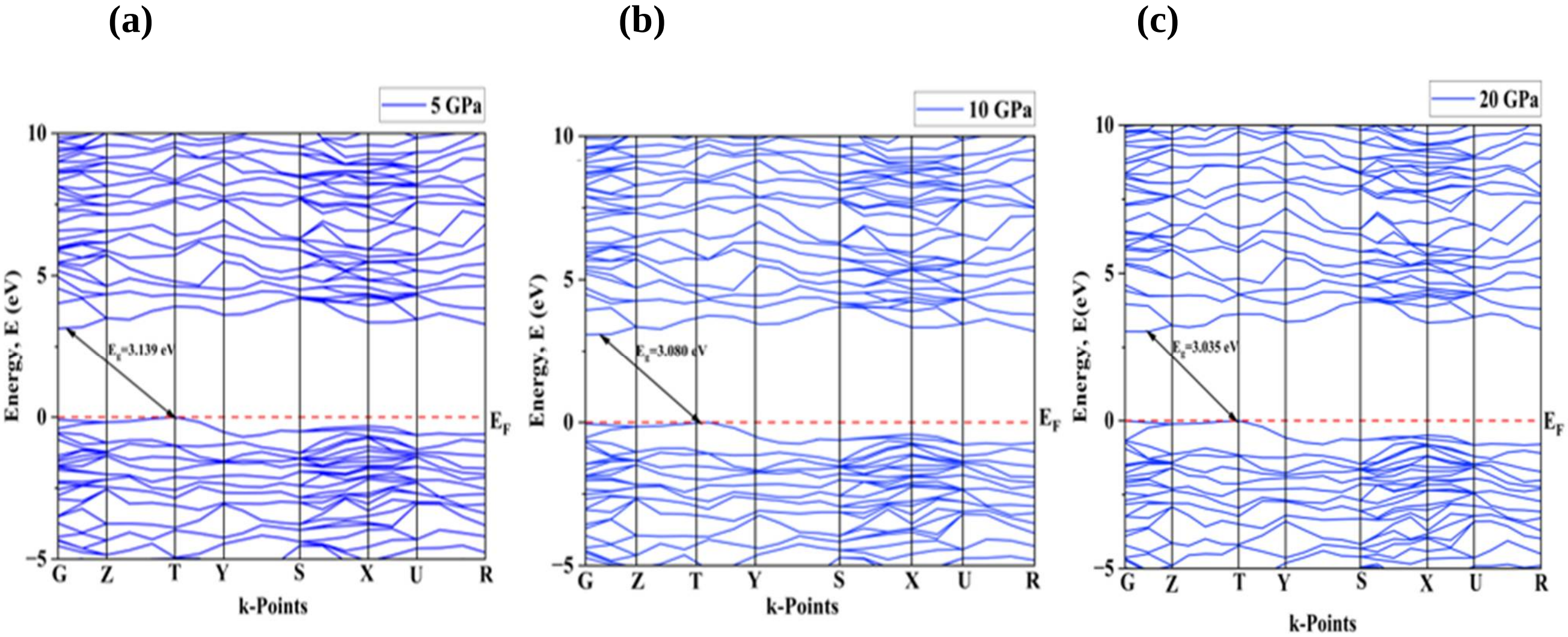

**Figure 21:** The calculated electronic band structures of the $B_6Se$ compound (a) 5 GPa, (b) 10 GPa and (c) 20 GPa along several high symmetry directions of the Brillouin zone in the ground state.

The position of the Fermi level with respect to the top of the valence band and bottom of the conduction band indicate that the conductivity of these two compounds should be dominated by holes in the intrinsic regime. The band gap (Eg) of $B_6X$ (X = S, Se) compounds as a function of hydrostatic pressure is illustrated in **Figure 22**. It shows that the band gap decreases with an increase in pressure. This tendency could be understood from the change in the electrical charge density. The unit-cell volume diminishes, and the bond length decreases under external pressure. As the total number of electrons is fixed in the system, both electronic charge density and electron overlap and interactions increase, which strengthens the bond strength and minimizes the band gap [92].

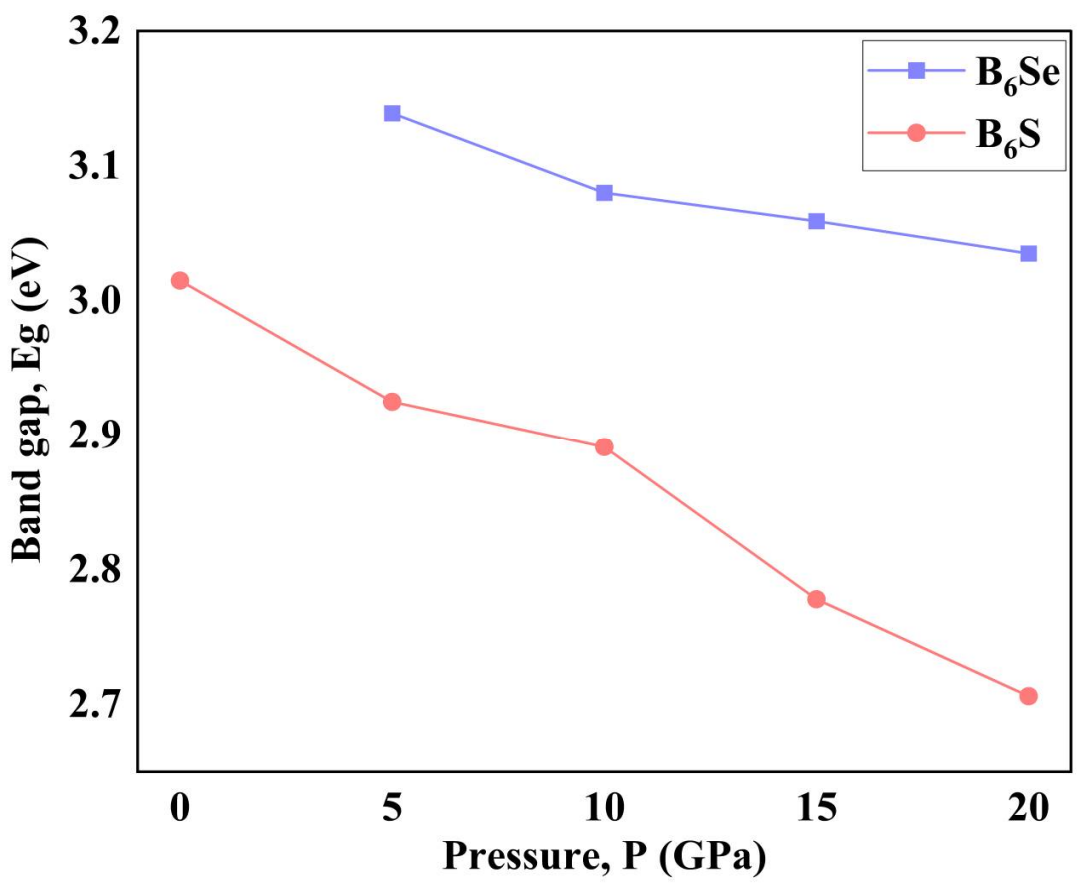

**Figure 22:** Band gap of $B_6S$ (red line) and $B_6Se$ (blue line) with various hydrostatic pressure.

**(b) Electronic Energy Density of States (EDOS)**

The electronic energy density of states (EDOS) of a crystalline material is important to understand all the charge and magneto-transport properties. It also gives information regarding the contribution of different atoms/orbitals to its optoelectronic capabilities, as well as how each atom contributes to bonding and anti-bonding states [95]. The DOS is the number of electronic states per unit energy per unit volume that are accessible for electrons to occupy. A high DOS for a given energy level indicates that there are many states accessible for occupancy. In this work, we calculated the total and partial density of states (TDOS and PDOS) of $B_6S$ and $B_6Se$ boron-rich chalcogenides for various pressures to acquire a better knowledge of their electronic properties. The calculated TDOS and PDOS are depicted in **Figures 23** and **24** for $B_6S$ at 0, 10, and 20 GPa and for $B_6Se$ at 5, 10 and 20 GPa, respectively, with the Fermi level ($E_F$) set at zero energy and indicated by the vertical dashed line. The TDOS values are nearly zero at the Fermi level for both compounds. It is seen from **Figure 23** that $B_6S$ exhibits a clear band gap in the TDOS profile, confirming its semiconducting nature [94], where the valence band is primarily contributed by the hybridization of B 2p and S 3p orbitals near $E_F$, while the B 2s and S 3s states dominate the lower energy region around -10 to -5 eV. Similar features are observed in $B_6Se$ as illustrated in **Figure 24**, a large band gap persists in the TDOS, with the bands near $E_F$ arising mainly from the strong hybridization between B 2p and Se 4p orbitals, whereas the Se 4s states are localized at deeper energies [93]. The PDOS analysis indicates that the s orbitals are situated in the low-energy valence region, but the hybridization of p orbitals enhances covalent bonding. The conduction bands are formed also by the same orbitals. The effect of pressure on the TDOS and PDOS profiles are not strong.

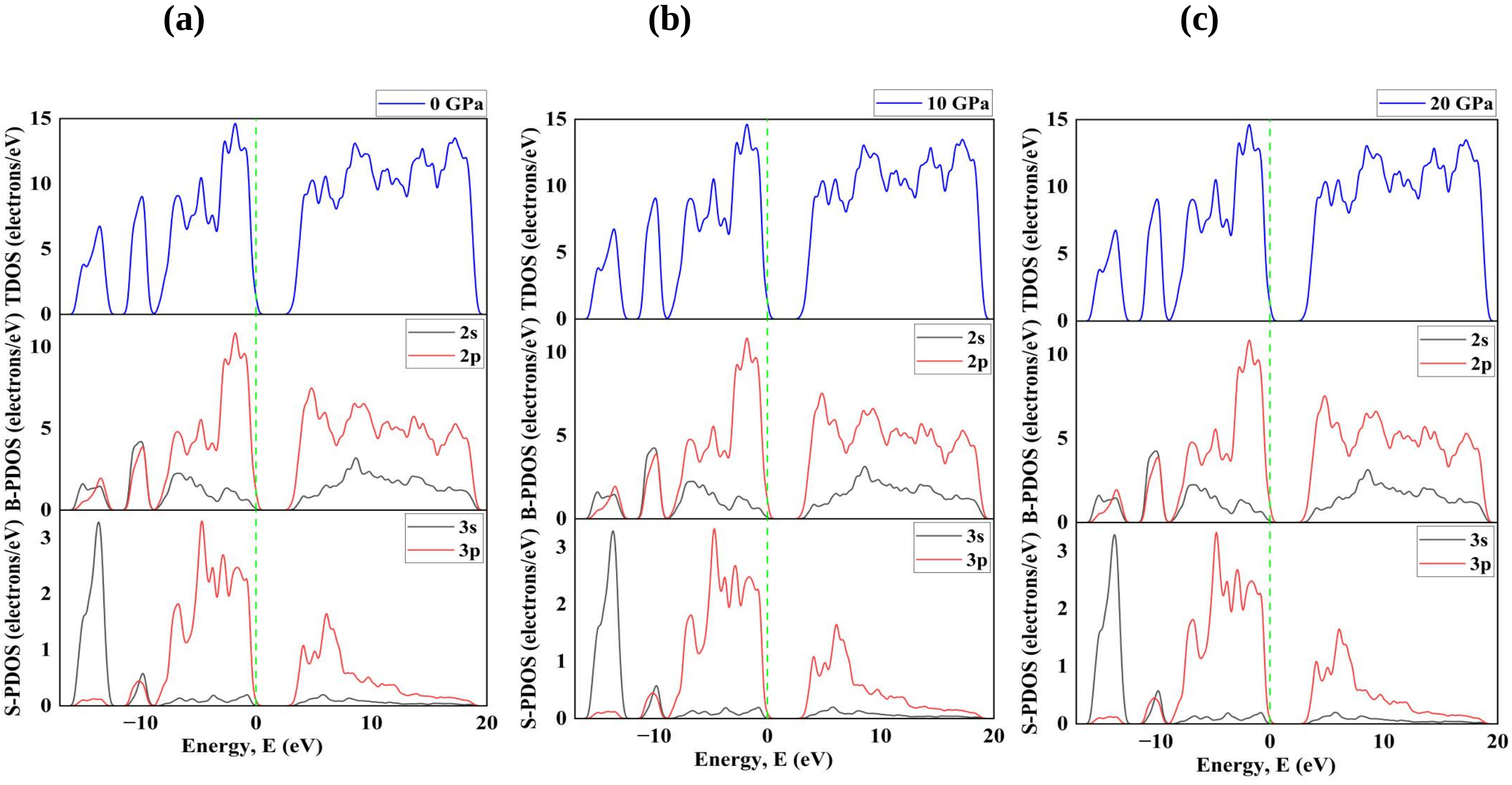

**Figure 23:** The total and partial density of states (TDOS and PDOS, respectively) plots of $B_6S$ for (a) 0 GPa, (b) 10 GPa and (c) 20 GPa pressure as a function of energy.

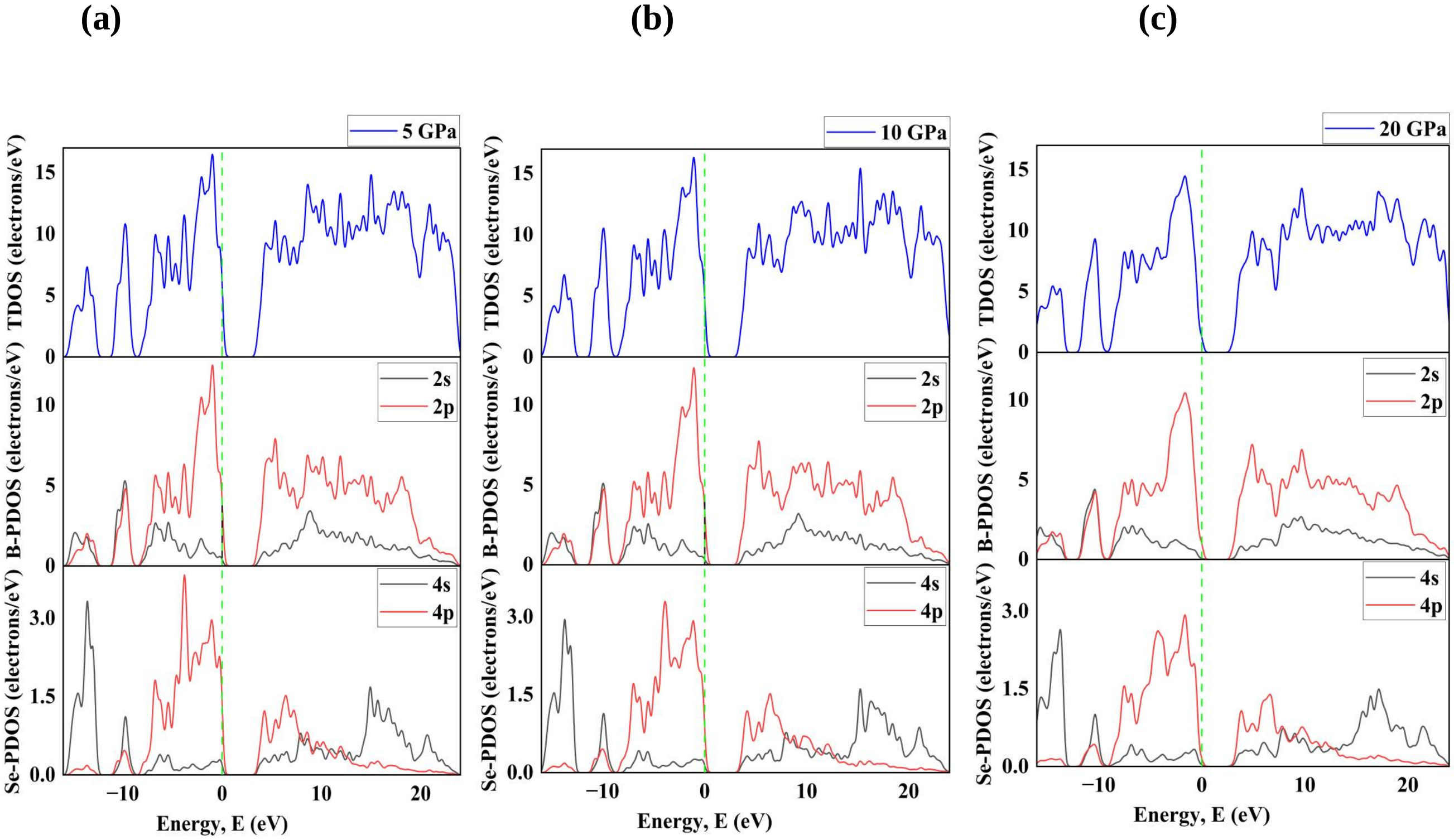

**Figure 24:** The total and partial density of states (TDOS and PDOS, respectively) plots of B6Se for (a) 5 GPa, (b) 10 GPa and (c) 20 GPa pressure as a function of energy.

### 3.8 Phonon Dynamics

The computation of phonon dispersion spectra (PDS) and phonon density of states (PHDOS) in crystalline materials has become a crucial aspect of materials research. These analyses are highly informative, as they provide essential insights into dynamic lattice stability or instability, phase transitions, and the vibrational contributions to heat conduction, thermal expansion, superconducting transition temperature, Helmholtz free energy, and heat capacity [102-104]. The generalized gradient approximation (GGA) with the Perdew–Burke–Ernzerhof (PBE) functional was employed to calculate the phonon dispersion spectra along the Γ–Z–T–Y–S–X–U–R paths within the Brillouin zone (BZ) and the phonon density of states (DOS) for the $B_6S$ and $B_6Se$ compounds. These calculations were carried out using the finite displacement method based on density functional perturbation theory (DFPT) [99,100].

The dynamical stability of a material is a key criterion for evaluating its suitability under varying mechanical stresses over time. **Figures 25** and **26** present the calculated phonon dispersion curves and phonon DOS along high-symmetry directions within the BZ for the $B_6S$ and $B_6Se$ compounds.

The absence of negative phonon frequencies across all pressures indicates the lack of soft phonon modes, confirming that both $B_6S$ and $B_6Se$ are dynamically stable under all pressures considered.

The acoustic modes consist of one longitudinal and two transverse branches, arising from the coherent vibration of atoms around their equilibrium positions. In contrast, the optical modes originate from out-of-phase oscillations between atoms, where one moves in one direction and the other in the opposite. Acoustic phonons are primarily responsible for sound propagation in crystals and are closely related to lattice rigidity. The three acoustic modes occur at lower frequencies, while the optical modes appear at higher frequencies, with partial overlap observed for $B_6S$ and $B_6Se$ across all pressures. Typically, low-frequency acoustic modes are associated with the vibrations of heavier atoms, whereas high-frequency optical modes result from the vibrations of lighter atoms. Notably, neither $B_6S$ nor $B_6Se$ exhibits a phonon band gap. In both compounds, the optical branches extend to considerably higher energies. To gain deeper insight into the lattice dynamics, the total and atomic partial PHDOS of $B_6S$ and $B_6Se$ were analyzed, in conjunction with the phonon dispersion curves, to identify the phonon bands and their corresponding density of states. For $B_6S$, prominent PHDOS peaks are observed at 29.99 and 33.71 THz for 0, and 20 GPa, respectively. Similarly, $B_6Se$ exhibits strong peaks at 29.42, and 32.22 THz for 5, and 20 GPa, respectively. The flatness of phonon bands leads to pronounced peaks in the PHDOS, while highly dispersive bands reduce the peak intensities. The convergence of the upper and lower phonon branches indicates the absence of a phononic band gap between the acoustic and optical modes in the PHDOS of both $B_6S$ and $B_6Se$ compounds.

**(a)**

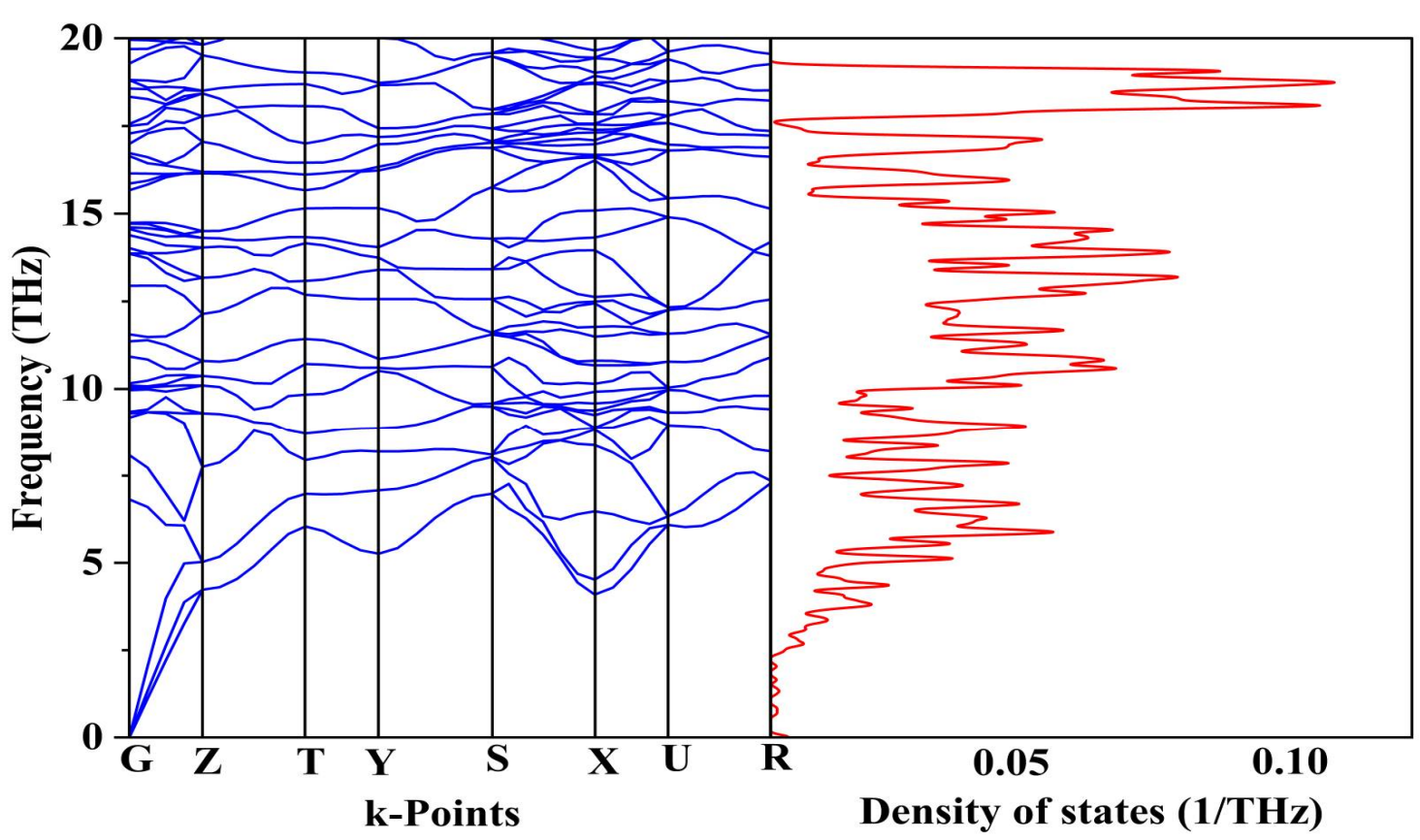

**(b)**

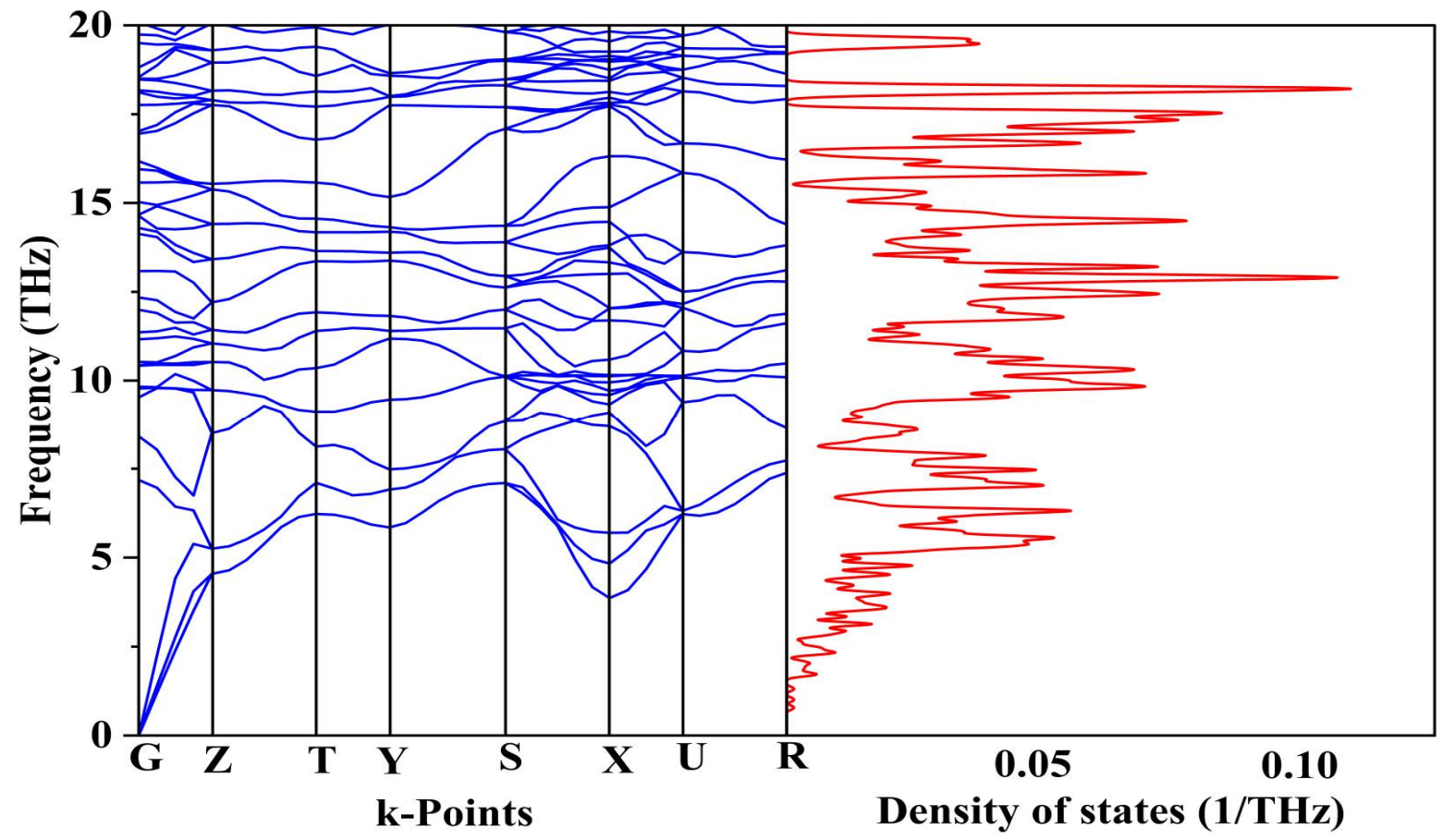

**Figure 25:** Phonon dispersion spectra and total phonon DOS of $B_6S$ compound (a) at 0 GPa and (b) at 20 GPa pressure.

**(a)**

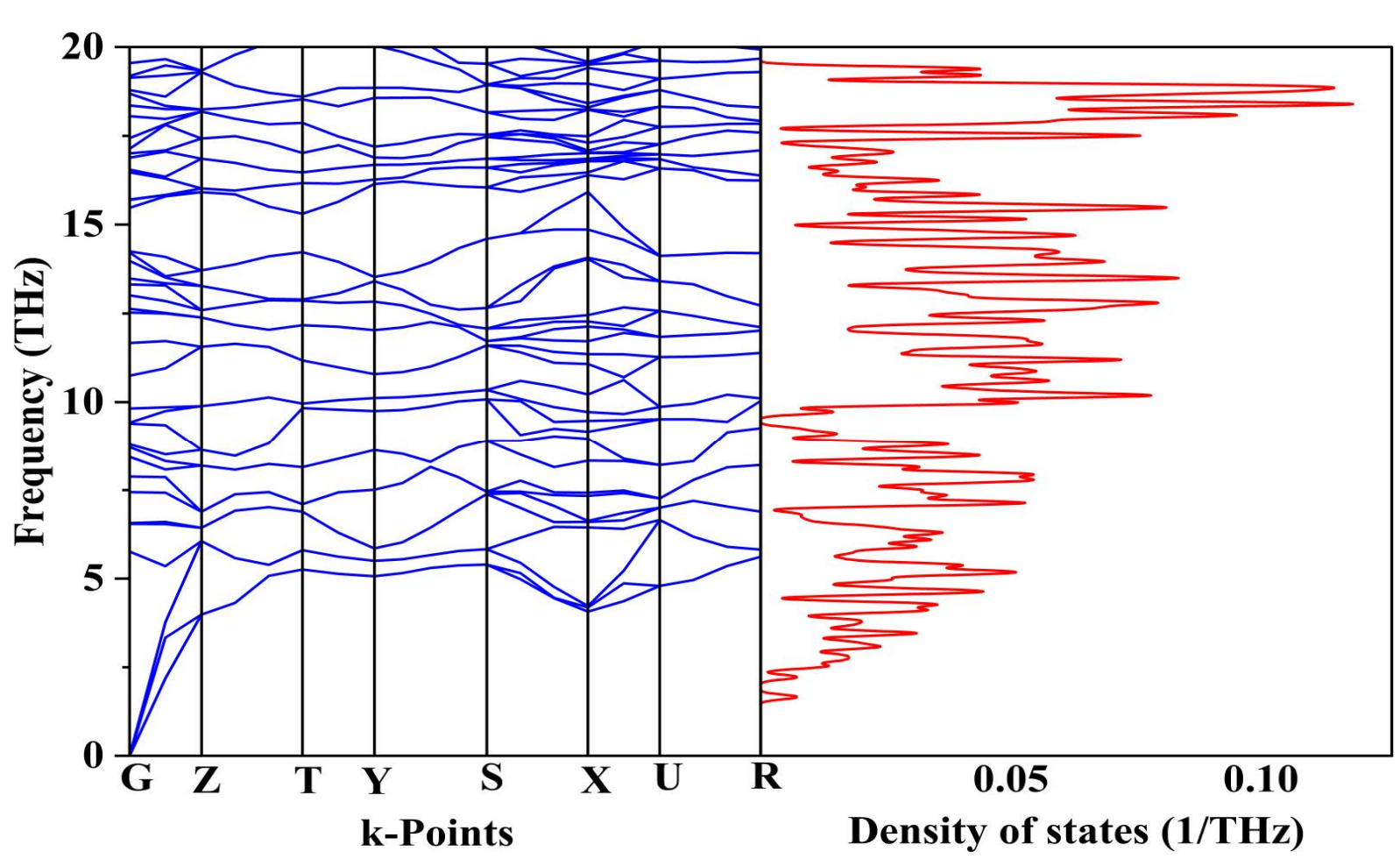

**(b)**

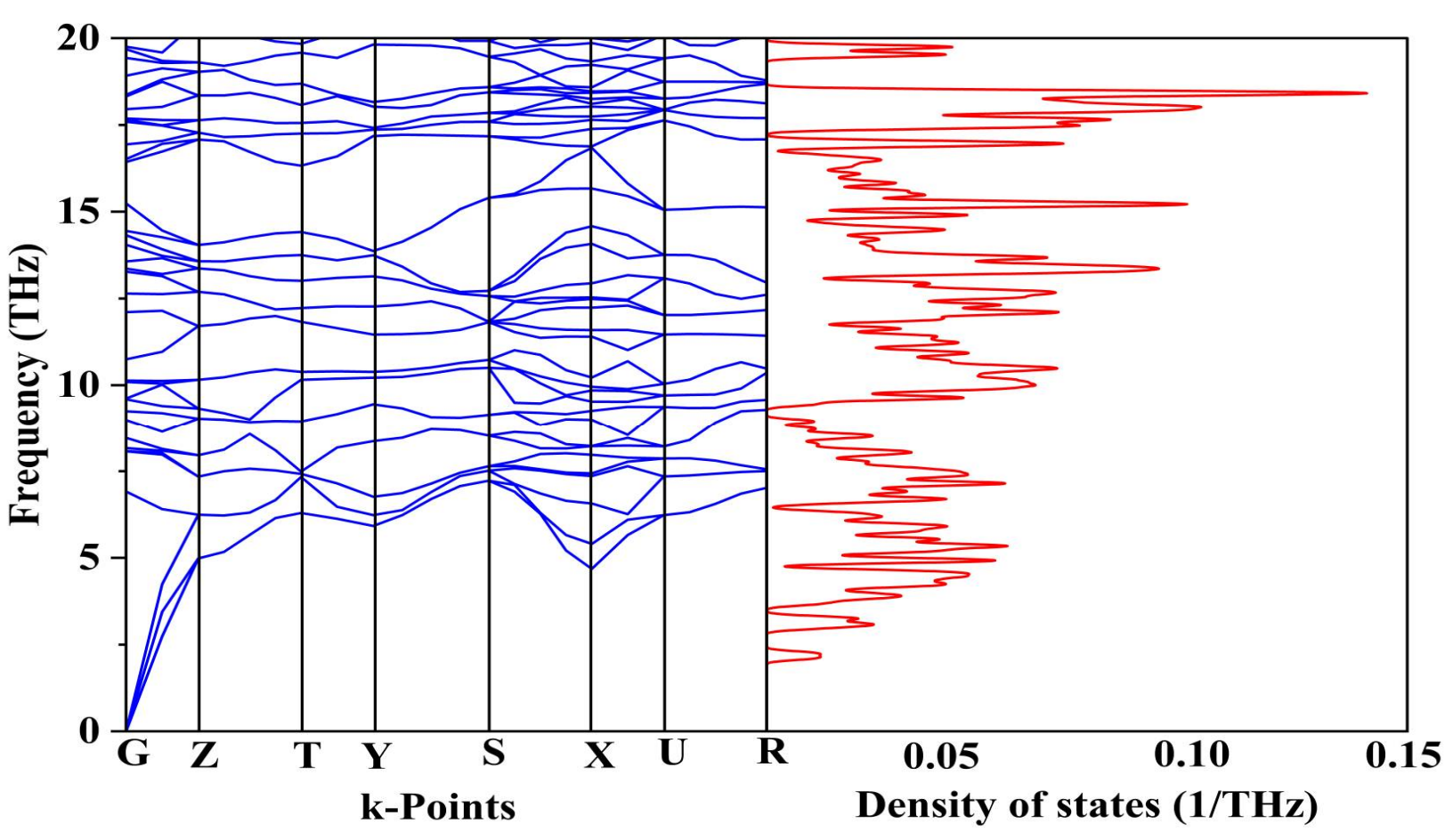

**Figure 26:** Phonon dispersion spectra and total phonon DOS of $B_6Se$ compound (a) at 5 GPa and (b) at 20 GPa pressure.

Relatively small PHDOS for low frequency branches are responsible for low thermal conductivity of the compounds under study.

## 4. Conclusions

A comprehensive first-principles investigation of orthorhombic $B_6S$ and $B_6Se$ under pressures ranging from 0-20 GPa reveals that both compounds remain structurally, mechanically, and dynamically stable within this pressure limit. Structural optimization shows continuous compression of lattice parameters and unit cell volume with pressure, where the b-axis is comparatively more compressible than the a- and c-axes. Elastic constant analysis confirms mechanical stability under Born-Huang criteria, while anisotropy in elastic response is evident from directional dependences of Young’s modulus, shear modulus, and compressibility. Both materials exhibit brittle nature (from B/G ratio, Poisson’s ratio, and Cauchy pressure) and increasing covalent character with pressure. Debye temperature and related thermal parameters indicate hard, thermally stable materials, while Grüneisen parameter suggests increasing anharmonicity under compression. The thermal properties of both $B_6S$ and $B_6Se$ are indicative of high levels of thermal stability and excellent characteristics for TBC applications. It has been shown that pressure can be used to tune the thermophysical properties of $B_6S$ and $B_6Se$ to customize the compounds for various engineering and structural applications.

Electronic band structure calculations show that both $B_6S$ and $B_6Se$ are wide band gap semiconductors, with band edges primarily governed by boron 2p states and minor contributions from S/Se p-orbitals. Pressure has negligible effect on band gap and density of states. Optical analysis confirms transparency in the visible region and strong absorption in the UV region, with moderately anisotropic optical response. The refractive index and dielectric function show polarization dependence, while energy loss spectra indicate plasmon peaks around 23-26 eV depending on direction. Mulliken population and electron density difference analyses confirm mixed ionic-covalent bonding with increasing covalency under pressure.

Overall, $B_6S$ and $B_6Se$ are mechanically stable, hard, brittle, and elastically anisotropic wide band gap semiconductors that maintain robust structural integrity under pressure up to 20 GPa. Their optical transparency in the visible region combined with strong UV absorption and stable electronic properties under compression suggests potential suitability for high-pressure optoelectronic and photonic applications.

## Acknowledgements

S.H.N. acknowledges the research grant (1151/5/52/RU/Science-07/19-20) from the Faculty of Science, University of Rajshahi, Bangladesh, which partly supported this work.

## Author Contributions

**Sourav Kumar Sutradhar:** Investigation, Data curation, Formal analysis, Visualization, Writing - original draft. **Tanvir Khan:** Methodology, Software, Writing - original draft. **Tauhidur Rahman:** Methodology, Formal analysis, Writing - original draft. **Sraboni Saha Moly:** Methodology, Formal analysis, Writing - original draft. **Mst. Maskura Khatun:** Methodology, Software, Writing - original draft. **S.H. Naqib:** Supervision, Methodology, Formal Analysis, Writing- Reviewing and Editing.

## Conflict of Interest

The authors declare no competing interests.

## Data Availability

The data sets generated and/or analyzed in this study are available from the corresponding author upon reasonable request.